\documentclass[pr,onecolumn,superscriptaddress]{revtex4}
\usepackage{amsmath}
\usepackage{amssymb}
\usepackage{amsthm}
\usepackage{amsfonts}
\usepackage{listings}
\usepackage{diagbox}
\usepackage{enumerate}
\usepackage{latexsym}
\usepackage{color}
\usepackage{xcolor}
\usepackage{bm}
\usepackage{hyperref}
\hypersetup{
 pdfnewwindow=true, colorlinks=true,
 linkcolor=blue, anchorcolor=blue,
 citecolor=blue, filecolor=blue,
 menucolor=blue, urlcolor=blue}

\newcommand{\beginsupplement}{%
        \setcounter{table}{0}
        \renewcommand{\thetable}{S\arabic{table}}%
        \setcounter{figure}{0}
        \renewcommand{\thefigure}{S\arabic{figure}}%
     }

\usepackage{listing}

\usepackage{psfrag}

\usepackage{bm}
\usepackage{graphicx}
\usepackage[FIGTOPCAP]{subfigure}
\usepackage[flushleft]{caption2}

\RequirePackage[normalem]{ulem} 
\RequirePackage{color}\definecolor{RED}{rgb}{1,0,0}\definecolor{BLUE}{rgb}{0,0,1} 

\newcommand{\braket}[2]{\left\langle #1 | #2 \right\rangle}
\newcommand{\bra}[1]{\left\langle#1\right|}
\newcommand{\ket}[1]{\left|#1\right\rangle}

\newcommand{\up}{\uparrow}
\newcommand{\dw}{\downarrow}

\newcommand{\bk}{{\bf k}}
\newcommand{\bv}{{\bf v}}
\newcommand{\br}{{\bf r}}
\newcommand{\bt}{{\bf t}}
\newcommand{\bg}{{\bf g}}
\newcommand{\bG}{{\bf G}}
\newcommand{\bL}{{\bf L}}

\begin{document}

\def\pgma{{\ttfamily irvsp}}
\def\pgmb{{\ttfamily ir2tb}}
\def\ie{{\it i.e.},\ }
\def\eg{{\it e.g.}\ }
\def\ea{{\it et al.}}
\def\et{{\it etc.}}
\input{epsf}

\tolerance 10000

\newcommand{\vk}{{\bf k}}

\draft

\title{{\ttfamily Irvsp}: to obtain irreducible representations of electronic states in the VASP}

\author{Jiacheng Gao}
\affiliation{Beijing National Laboratory for Condensed Matter Physics,
and Institute of Physics, Chinese Academy of Sciences, Beijing 100190, China}
\affiliation{University of Chinese Academy of Sciences, Beijing 100049, China}

\author{Quansheng Wu}
\affiliation{Institute of Physics, \'{E}cole Polytechnique F\'{e}d\'{e}rale de Lausanne, CH-1015 Lausanne, Switzerland}
\affiliation{National Centre for Computational Design and Discovery of Novel Materials MARVEL, Ecole Polytechnique F\'{e}d\'{e}rale de Lausanne (EPFL), CH-1015 Lausanne, Switzerland}

\author{Clas Persson}
\affiliation{Centre for Materials Science and Nanotechnology, Department of Physics, 
University of Oslo, P.O. Box 1048 Blindern, NO-0316 Oslo, Norway}
\affiliation{Department of Materials Science and Engineering, 
KTH Royal Institute of Technology, Stockholm, SE-100 44, Sweden}

\author{Zhijun Wang}
\email{zjwang11@hotmail.com}
\altaffiliation{\\$~$ The codes are available in the public repository: \url{https://github.com/zjwang11/irvsp/}.}
\affiliation{Beijing National Laboratory for Condensed Matter Physics,
and Institute of Physics, Chinese Academy of Sciences, Beijing 100190, China}
\affiliation{University of Chinese Academy of Sciences, Beijing 100049, China}


\begin{abstract}
We present an open-source program \href{https://github.com/zjwang11/irvsp/blob/master/src\_irvsp\_v2.tar.gz}{\ttfamily irvsp}, to compute irreducible representations of electronic states for all 230 space groups with an interface to the Vienna \emph{ab-initio} Simulation Package. This code is fed with plane-wave-based wavefunctions (\eg WAVECAR) and space group operators (listed in OUTCAR), which are generated by the VASP package.
This program computes the traces of matrix presentations and determines the corresponding irreducible representations for all energy bands and all the $k$-points in the three-dimensional Brillouin zone. It also works with spin-orbit coupling (SOC), \ie for double groups. It is in particular useful to analyze energy bands, their connectivities, and band topology, after the establishment of the theory of topological quantum chemistry. Accordingly, the associated library -- \href{https://github.com/zjwang11/irvsp/blob/master/lib\_irrep\_bcs.tar.gz}{\ttfamily irrep\_bcs.a} -- is developed, which can be easily linked to by other \emph{ab-initio} packages.  In addition, the program has been extended to orthogonal tight-binding (TB) Hamiltonians, \eg electronic or phononic TB Hamiltonians. A sister program \href{https://github.com/zjwang11/irvsp/blob/master/src\_ir2tb\_v2.tar.gz}{\ttfamily ir2tb} is presented as well.\\
\textbf{Program summary} \\
{\it Program title:} {\ttfamily irvsp} \\
{\it Program Files doi:} \url{https://doi.org/10.17632/y9ds5nnm2f.1} \\
{\it Licensing provisions:} GNU Lesser General Public License, \url{https://www.gnu.org/licenses/lgpl-3.0.html} \\
{\it Distribution format:} {\ttfamily tar.gz} \\
{\it Programming language:} Fortran 90/77 \\
{\it Computer:} Any architecture with a Fortran 90 complier \\
{\it RAM:} 20 MB \\
{\it Nature of problem:}  
Determining irreducible representations for all energy bands and all the $k$-points in 230 space groups. It is in particular useful to analyze energy bands, their connectivities, and band topology. \\
{\it Solution method:} 
By computing the traces of matrix presentations of space group operators for the eigen-wavefunctions at a certain $k$-point in a given space group, one can determine irreducible representations for them. \\
{\it Running time:} It takes less than 1 minute for the calculation of Bismuth. 
\end{abstract}

\maketitle
\section{Introduction}
Topological states have been intensively studied in the past decades~\cite{bernevig2006quantumsci,konig2007quantum,kane2005quantum,bernevig2006quantumprl,kane2010colloquium,qi2011topological,schindler2018higher,wan2011topological,wieder2018wallpaper}. During the period, lots of materials have theoretically been proposed to be topological insulators and topological semimetals, based on calculations within the density-functional theory (DFT)~\cite{tang2018towards,zhang2019catalogue,vergniory2019complete,yuanfeng2019,ligang2017,tase32018,hfrup2019}. Many of them are verified in experiments, and substantially intrigue much interest in theories and experiments, such as three-dimensional (3D) topological insulator Bi$_2$Se$_3$~\cite{zhang2009topological,xia2009observation,chen2009experimental}, Dirac semimetals Na$_3$Bi~\cite{wang2012dirac,liu2014discovery} and Cd$_3$As$_2$~\cite{PhysRevB.88.125427,liu2014stable}, Weyl semimetal TaAs~\cite{weng2015weyl,huang2015weyl,lv2015experimental,xu2015discovery}, topological crystalline insulator SnTe~\cite{hsieh2012topological,tanaka2012experimental} and hourglass material KHgSb~\cite{wang2016hourglass,ma2017experimental} \ea~To some extent, these topological electron bands are related to a band-inversion feature. Explicitly, there can be a band inversion between different irreducible representations (IRs) of the little groups at $k$-points in the 3D Brillouin zone (BZ)~\cite{zhu2012band}. In the situation of Dirac semimetals or nodal-line semimetals, the band inversion may happen on a high-symmetry line or in a high-symmetry plane.

Very recently, new insights about band theory have been used to classify all the nontrivial electron band topologies compatible with a given crystal structure~\cite{Ashvin2017,song2018,slager2017,bradlyn2017topological}. In particular, based on the theory of topological quantum chemistry (TQC)~\cite{bradlyn2017topological,cano2018building,vergniory2017graph,cano2018topology}, 
the topology of a set of isolated electron bands is relied on IRs at the maximal high-symmetry $k$-points (HSK), as the compatibility relations are obtained in Ref.~\cite{elcoro2017double}, and open accessible on the Bilbao Crystallographic Server (BCS)~\cite{aroyo2011crystallography,stokes2013tabulation}. The set of maximal HSK points can be found by using the BCS. The determination of the IRs of electron bands at maximal HSK points is of great interest, for which the program -- {\ttfamily vasp2trace} -- was developed~\cite{vergniory2019complete}. However, it is not suitable for any non-maximal HSK points.

Generally speaking, in order to obtain the IRs for electron energy bands in crystals, two ingredients are necessary:  a) wave-functions (WFs) at $k$-points and b) character tables (CRTs) for $k$-little groups. Different versions of the codes can be developed based on the different types of the WFs and conventions of the CRTs.
The program {\ttfamily irrep} in the WIEN2k package~\cite{blaha2001wien2k,persson1999} is a precursor in determining the IRs,
which is based on the plane-wave (PW) basis (the part of the WFs outside muffin-tin spheres) and the CRTs of 32 point groups (PNGs). 
There is an advantage of describing the IRs in terms of the more well-known PNG symmetries; however, the disadvantage is that in many cases $k$-points on the BZ surface cannot be classified with PNGs for nonsymmorphic crystals.
In this paper, the program -- {\ttfamily irvsp} -- is developed based on the CRTs on the BCS. 
It originates from the WIEN2k {\ttfamily irrep} code~\cite{blaha2001wien2k,persson1999} that considers both single and double groups, analyses of time-reversal symmetry, and handles accidental degeneracies. The present code inherits those features but it has been extended to also be able to determine IRs of those special $k$-points for nonsymmorphic crystals. Hence, the code labels the IRs according to the convention of the BCS notation~\cite{stokes2013tabulation} for 230 space groups (SGs). In fact, it works for 1651 magnetic space groups (MSGs), once the space group number of the unitary part of MSGs is correctly given.
In addition, Wannier-based tight-binding (TB) models are widely used to study the topological properties of real materials, including topological surface states and symmetry indicators.
To get the band representations and check the topology of these models, a sister program -- {\ttfamily ir2tb} -- is developed to interface with orthogonal TB Hamiltonians, \eg electronic or phononic TB Hamiltonians.

\ \\ 

This paper is organized as follows. In Section~\ref{sec:2}, we present some basic derivations to compute the traces of matrix presentations (MPs) in different bases. In Section~\ref{sec:3}, we introduce the general process of the code. In Section~\ref{sec:4}, we introduce the capabilities of this package. In Section~\ref{sec:5}, we introduce the installation and basic usages. In Section~\ref{sec:6}, we introduce some examples in order to show how to use \pgma\ to determine the IRs and further explore the topology.

\section{Methods}
\label{sec:2}
Space-group operations (SGOs), ${\cal O}_s=\{R_s|\bv_s\}$, are consist of two parts: a rotation part $R_s$ and a translation part $\bv_s$. The product of two operations is defined as $\{R_s|\bv_s\}\{R_t|\bv_t\}=\{R_sR_t|R_s\bv_t+\bv_s\}$. An operator acting on a scalar function in real space is expressed by ${\cal O}_sf(\br)= f({\cal O}_s^{-1}\br)=f( R_s^{-1}\br-R_s^{-1}\bv_s)$ (There is a typo in Section A of the supplementary information of Ref.~\cite{vergniory2019complete}).
The MPs, $O_{s}^{mn}$, can be computed in the basis of the Bloch wavefunctions $\ket{\psi_{n\bf k}}$: $O_{s}^{mn}=\braket{\psi_{m\bf k}}{{\cal O}_{s}|\psi_{n\bf k}}$. The traces of the obtained MPs are the characters, and they are essential to determine the corresponding IRs of the little group (LG) of $\bk$.  The LG of $\bk$ [$LG(k)$] is defined as a set of SGOs: 
\begin{equation}
LG(k):~\{{\cal O}_s|R_s\bk=\bk+\bG\}, \text{ with } \bG=l\bg_1+m\bg_2+n\bg_3,~l,m,n\in \mathbb N
\label{eq:lgk}
\end{equation}
Here, $\bG$ could be any integer reciprocal lattice translation ($\bg_1, \bg_2, \bg_3$ are primitive reciprocal lattice vectors). The traces of MPs of SGOs are defined as:
\begin{equation}
 \text{Tr}[{\cal O}_s]=\sum_{n}O_s^{nn}\text { with } O_s^{nn}=\braket{\psi_{n\bf k}}{{\cal O}_s|\psi_{n\bf k}},~ {\cal O}_s\in LG(k).
\label{eq:mat}
\end{equation}
Here, the WFs have to be normalized (\ie $\braket{\psi_{n\bf k}}{\psi_{n\bf k}}=1$).

Under different bases, the WFs can be expressed in different ways, and the derivations of Eq.~(\ref{eq:mat}) are different. Here, we have derived the expressions in two bases: i) PW basis and ii) TB basis. In what follows, symbols in the bold text are vectors, and common braket notations are employed:
\begin{eqnarray*}
\braket{\br}{A}&\equiv&  A(\br)\\
\braket{A}{B} &\equiv& \int d\br A^*(\br) B(\br) \\
\braket{\br}{\bk}&\equiv& e^{i\bk\cdot \br} 
\end{eqnarray*}
To be convenient, we present the derivations in the cases without the spin degree of freedom. However, the derivations can be easily extended to the cases including SOC, by substituting $R_s\otimes SU_s(2)$ for $R_s$, where the bases are doubled by the direct product: $ \{{\rm PW/TB}~basis\} \otimes \{\ket{\up},\ket{\dw}\}$. In fact, the code works for both single and double groups.

\subsection{Plane-wave basis}
In the PW basis, wavefunctions/eigenstates are expressed in the basis of plane waves:
\begin{align}
&\psi_{n\bf k}(\br)= \sum_j C_{\bk,j} e^{i(\bk+\bG_j)\cdot \br} 
\text{ with }\braket{\bk+\bG_i}{\bk+\bG_j}=\delta_{ij} 
\label{eq:3}
\end{align}
The coefficients ($C_{\bk,j}$) are usually computed in the \emph{ab-initio} calculations and output by the DFT package (\eg VASP, PWscf, \et).
The SGOs acting on WFs are derived as:
\begin{eqnarray*}
{\cal O}_s\psi_{n\bf k}(\br)&=&\sum_j C_{\bk,j} e^{i(\bk+\bG_j)\cdot (R^{-1}_s\br-R^{-1}_s\bv_s)} \\
                                      &=&\sum_j C_{\bk,j} e^{iR_s(\bk+\bG_j)\cdot (\br-\bv_s)}\\
                                      &=&\sum_j C_{\bk,j} e^{i(\bk+\bG_{j'})\cdot (\br-\bv_s)}\text{ with } \bk+\bG_{j'}\equiv R_s(\bk+\bG_{j})\\
                                      &=&e^{-i\bk\cdot \bv_s}\sum_j C_{\bk,j} e^{-i\bG_{j'}\cdot \bv_s}e^{i(\bk+\bG_{j'})\cdot \br}\text{ with } \bG_{j'}\equiv R_s(\bk+\bG_{j})-\bk \\
\end{eqnarray*}
Then, Eq.~(\ref{eq:mat}) can be written as:
\begin{eqnarray}
\braket{\psi_{n\bf k}}{{\cal O}_s|\psi_{n\bf k}}&=&e^{-i\bk\cdot \bv_s} \sum_j C^*_{\bk,j'} C_{\bk,j} e^{-i\bG_{j'}\cdot \bv_s} \text{ with } \bG_{j'}\equiv R_s(\bk+\bG_{j})-\bk 
\label{eq:pw}
\end{eqnarray}
The program \pgma~is developed based on the above derivations with the interface to VASP~\cite{kresse1996software}. In addition, the library of the code is developed (see details in Appendix~\ref{sup:6}), which can be
downloaded from the public code archive: \url{https://github.com/zjwang11/irvsp/blob/master/lib\_irrep\_bcs.tar.gz}. The library -- {\ttfamily irrep\_bcs.a} -- can be easily linked to by other \emph{ab-initio} packages, once a proper interface is made.

\subsection{Orthogonal tight-binding basis}
In a TB Hamiltonian, WFs are expressed in the basis of exponentially localized orthogonal orbitals: $\ket{\bf{0},\mu\alpha}\equiv\phi_{\alpha}^\mu(\br)\equiv \phi_\alpha(\br-\tau_\mu)$ and $\ket{\bL_j,\mu\alpha}\equiv \phi_\alpha(\br-\bL_j-\tau_\mu)$, where $\mu$ label the atoms,
 $\alpha$ label the orbitals, $\bL_j$ label the lattice vectors in 3D crystals, and $\tau_\mu$ label the positions of atoms in the home unit cell.
At a given $k$-point, WFs are given as:
\begin{align}
&\psi_{n\bf k}(\br)= \sum_{\mu\alpha} C^n_{\mu\alpha,\bk} \phi_{\alpha \bk}^\mu(\br) \text{ where~$n$~is a band index},  \\
&\phi_{\alpha \bk}^\mu(\br)=\sum_j \phi_\alpha(\br-\tau_\mu-\bL_j)e^{i \mathbf{k}\cdot (\bL_j+\tau_\mu)}, \braket{\phi_{\beta\bk}^{\mu'}}{\phi_{\alpha\bk}^{\mu}}=\delta_{\mu\mu'}\delta_{\alpha\beta}\label{eq:fft}
\end{align}
The states $\phi_{\alpha \bk}^\mu(\br)$ are the Fourier transformations of the local orbitals $\phi_{\alpha}^\mu(\br)$, as shown in Eq. (\ref{eq:fft}).
The coefficients are obtained as the eigenvectors of the TB Hamiltonian:
$H_{\mu'\beta,\mu\alpha}(\bk)=\sum_j e^{i\bk\cdot (\bL_j+\tau_\mu-\tau_{\mu'})}\braket{{\bf{0}},\mu'\beta}{\hat H|\bL_j, \mu\alpha}$.
The rotational symmetries $R_s$ acting on the local orbitals [$\phi_{\alpha}(\br)$] at the $\mu$ site are given as:
\begin{align}
\widehat{R_s\phi}_{\alpha}(\br)\equiv R_s\phi_{\alpha}(\br)=\sum_\beta\phi_{\beta}(\br) D^{s,\mu}_{\beta\alpha}
\label{eq:Drep}
\end{align}
These $D$-matrices are explicitly given in Table \ref{tab:conv1} in Appendix~\ref{sup:2}. Under the basis of real spherical harmonic functions with different total angular momenta (integer $l$), these $D$-matrices are real.

The SGOs acting on the states $\phi_{\alpha \mathbf{k}}^{\mu}(\br)$ are given below:
\begin{eqnarray*}
{\cal O}_s \phi_{\alpha \mathbf{k}}^{\mu}(\br)&=&\phi_{\alpha \mathbf{k}}^{\mu}(R_s^{-1}\br-R_s^{-1}\bv_s) \notag \\
&=& \sum_j \phi_\alpha(R_s^{-1}\br-R_s^{-1}\bv_s-\tau_\mu-\bL_j)e^{i \mathbf{k}\cdot (\bL_j+\tau_\mu)} \notag \\
&=& \sum_j \phi_\alpha(R_s^{-1}[\br-\bv_s-R_s\tau_\mu-R_s\bL_j])e^{i \mathbf{k}\cdot (\bL_j+\tau_\mu)} \notag \\
&=& \sum_j \widehat{R_s\phi}_{\alpha}[\br-\bv_s-R_s\tau_\mu-R_s\bL_j]e^{i (R_s\mathbf{k})\cdot R_s(\bL_j+\tau_\mu)} \text{ with } \widehat{R_s\phi}_{\alpha}(\br)\equiv \sum_\beta\phi_{\beta}(\br) D^{s,\mu}_{\beta\alpha}\notag \\
&=& e^{-i(R_s\bk\cdot \bv_s)}\sum_j \widehat{R_s\phi}_{\alpha}[\br-(\bv_s+R_s\tau_\mu)-R_s\bL_j]e^{i (R_s\mathbf{k})\cdot [R_s\bL_j+(R_s\tau_\mu+\bv_s)]} \notag \\
&=& e^{-i(R_s\bk\cdot \bv_s)}\sum_j \widehat{R_s\phi}_{\alpha}[\br-(\tau_{\mu'}+\bL^i_0)-R_s\bL_j]e^{i (R_s\mathbf{k})\cdot [R_s\bL_j+(\tau_{\mu'}+\bL^i_0)]} \text{ using }\bv_s+R_s\tau_\mu=\bL^i_0+\tau_{\mu'} \notag\\
&=& e^{-i(R_s\bk\cdot \bv_s)}\sum_{j'} \widehat{R_s\phi}_{\alpha}[\br-\tau_{\mu'}-\bL_{j'}]e^{i (R_s\mathbf{k})\cdot [\bL_{j'}+\tau_{\mu'}]}\text{ with } \bL_{j'}=\bL^i_0 +R_s\bL_j \notag \\
&=& e^{-i(R_s\bk\cdot \bv_s)}\sum_\beta\phi_{\beta,R_s\mathbf{k}}^{\mu'}(r) D^{s,\mu}_{\beta\alpha}\notag \\
\end{eqnarray*}
Thus, Eq.~(\ref{eq:mat}) is written as:
\begin{eqnarray}
\braket{\psi_{n\bf k}}{{\cal O}_s|\psi_{n\bf k}}&=&e^{-i(R_s\bk\cdot \bv_s)} \sum_{\alpha\mu,\beta}  (C^{n}_{\mu'\beta})^* e^{i(R_s\bk-\bk)\cdot \tau_{\mu'}} D^{s,\mu}_{\beta\alpha} C^n_{\mu\alpha} 
 \text{ with } \bv_s+R_s\tau_\mu=\bL^i_0+\tau_{\mu'} 
\end{eqnarray}
In a matrix format, 
\begin{eqnarray}
&\braket{\psi_{n\bf k}}{{\cal O}_s|\psi_{n\bf k}}=e^{-i(R_s\bk\cdot \bv_s)}\left[\overline{C^{\dagger}V(R_s\bk-\bk)DC}\right]_{nn} 
\label{eq:9} \\
&\text{ with } \overline{V}(\bk)_{\mu'\beta,\mu\alpha}  = e^{i\bk\cdot {\tau_\mu}}\delta_{\mu\mu'}\delta_{\alpha\beta},~\overline{C}_{\mu\alpha,n}=C^n_{\mu\alpha},
\overline{D}_{\mu'\beta,\mu\alpha}=\begin{cases}
\begin{array}{cc}
D^{s,\mu}_{\beta\alpha} &\text{when }\bv_s+R_s\tau_\mu=\bL^i_0+\tau_{\mu'};\\
                   0   &\text{otherwise}.
\end{array}\end{cases}
\end{eqnarray}

Based on the above derivations, the code has been extended to the TB basis. The sister program is called \pgmb.
To run \pgmb, users must provide two files: {\ttfamily case\_hr.dat} and {\ttfamily tbbox.in}. The file called {\ttfamily case\_hr.dat}, containing the TB parameters, may be generated by the software Wannier90~\cite{wannier90rmp,mostofi2014updated} with symmetrization~\cite{WU2017,yue2018symmhr,gresch2018automated}, or generated by users with a toy TB model, or generated from Slater-Koster method~\cite{sk1954} or discretization of $k\cdot p$ model onto a lattice~\cite{kpmethod}. The other file {\ttfamily tbbox.in} is the master input file for \pgmb. It should be given consistently with the TB parameters in {\ttfamily case\_hr.dat}. 
The {\ttfamily tbbox.in} for Bi$_2$Se$_3$ is given in Appendix~\ref{sup:1}. In addition, the example of electronic TB Hamiltonian for Bi$_2$Se$_3$ and the example of phononic TB Hamiltonian for CoSi are included in the archive {\ttfamily src\_ir2tb\_v2.tar.gz}.

\begin{table}[!h]
  \captionstyle{centerlast}
  \caption{
A brief summary of key subroutines
  }
  \begin{tabular}{p{2.6cm}p{8cm}p{0.6cm}c}
  \hline
  \hline
  File & Description & &Input \\
  \hline
  {\ttfamily wave\_data.f90} & reading the coefficients $C_{{\bf k},j}$. && WAVECAR \\ 
  \hline
  {\ttfamily init.f90}  & reading lattice vectors and space group operators,  
   and setting up the $Z$ and $Z^{-1}$ matrices.  && OUTCAR \\ 
  \hline
 {\ttfamily kgroup.f90} & determining the $\bk$-little groups. & &\\ 
  \hline
 {\ttfamily nonsymm.f90} & retrieving the character tables from the BCS& & \\
  \hline
{\ttfamily chrct.f90} & computing the traces through Eq.~(\ref{eq:pw}), and determining the IRs& &\\ 
  \hline
  \hline
  \end{tabular}
  \label{tab:sum}
\end{table}

\section{General process of the code}
\label{sec:3}
In the main text, we are mainly focused on \pgma, which is based on the PW basis with an interface to the VASP package~\cite{kresse1996software}. The key subroutines are summarized in Table~\ref{tab:sum}. One can check more details for \pgmb~and {\ttfamily irrep\_bcs.a} in the Appendix. The program \pgmb~works for orthogonal TB Hamiltonians, \eg the electronic or phononic TB Hamiltonians. The library {\ttfamily irrep\_bcs.a} can be linked to by other DFT packages. 

\subsection{Wavefunctions at $k$-points}
In the VASP package, the all-electron wave-function is obtained by acting a linear operator $\cal T$ on the pseudo-wavefunction: $\ket{\psi_{n\bf k}}={\cal T}\ket{\tilde \psi_{n\bf k}}$. The linear operator can be written explicitly as: ${\cal T}={\bf 1}+\sum_i \left (\ket{\phi_i}-\ket{\tilde \phi_i}\right)\bra{p_i}$, where $\ket{\phi_i}$ ($\ket{\tilde \phi_i}$) is a set of all-electron (pseudo) partial waves around each atom and $\ket{p_i}$ is a set of corresponding projector functions on each atom within an augmentation region ($r<R_0$), where $R_0$ is the core part for each atom.  
The pseudo-wavefunction is expanded in the plane waves:
\begin{equation}
\tilde \psi_{n\bf k}(\vec r) \equiv
\braket{\vec r}{\tilde \psi_{n\bf k}}=\sum_{\bG_j}C_{n,{\bf k}+\bG_j}e^{i({\bf k}+\bG_j)\cdot \vec r} 
\end{equation}
where $\bG_j$ vectors are determined by the condition $\frac{\hbar^2}{2m_e}({\bf k}+\bG_j)^2<E_{cutoff}$ with a cutoff $E_{cutoff}$.
It is worthy noting that $\ket{\tilde{\psi}_{n\bk}}$ are sufficient for the calculations of the traces of MPs of SGOs. 

Since the pseudo-wavefunctions $\ket{\tilde \psi_{n\bk}}$ are usually not normalized, they have to be renormalized before their traces can be computed via Eq.~(\ref{eq:mat}).
The coefficients ($C_{\bk+\bG_j}$) are output in WAVECAR by VASP. 
In the program, they are read by the subroutine: {\ttfamily wave\_data.f90}.
In the SOC case, the $C_{\bk+\bG_j,\up}$ and $C_{\bk+\bG_j,\dw}$ are stored in the complex variables {\ttfamily coeffa(:)} and {\ttfamily coeffb(:)}. In the case without SOC, the $C_{\bk+\bG_j}$ are stored in {\ttfamily coeffa(:)}, while {\ttfamily coeffb(:)} are invalid (set to be zero).

\subsection{Space group operators of a 3D crystal}
\begin{figure}[htb]
  \raggedright
  \includegraphics[scale=0.45]{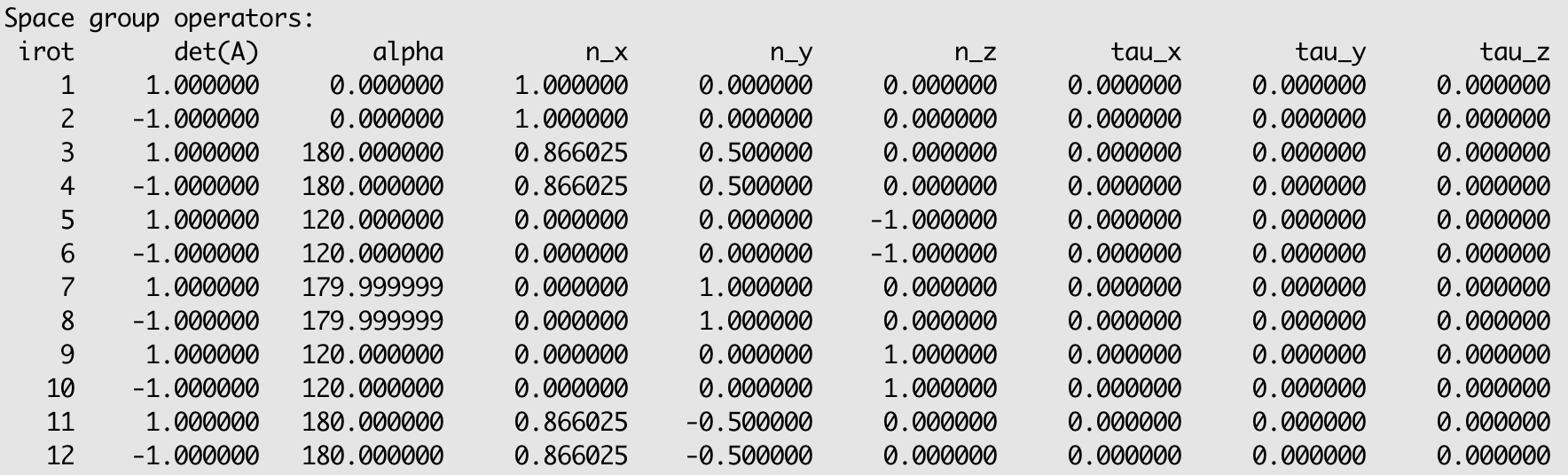}
  \caption{Screenshot of OUTCAR, showing the space group operators of Bi$_2$Se$_3$ generated by VASP.}
  \label{fig:bise_spo}
\end{figure}

\begin{figure}[htb]
  \raggedright
  \includegraphics[scale=0.45]{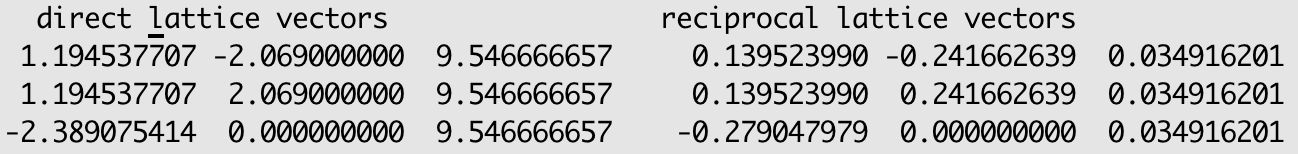}
  \caption{Screenshot of OUTCAR, showing the lattice vectors and reciprocal lattice vectors of Bi$_2$Se$_3$ which are used in VASP.}
  \label{fig:bise_lat}
\end{figure}

Instead of generating space group operators from a 3D crystal structure (\ie POSCAR), the program reads the SGOs directly from the standard output of VASP (\ie OUTCAR), which is done by the subroutine: {\ttfamily init.f90}. In other words, the SGOs are generated by the VASP package (\eg with {\ttfamily ISYM = 1} or {\ttfamily 2}  in INCAR for vasp.5.3.3), which are listed below the line of `Space group operators:' in OUTCAR. Fig.~\ref{fig:bise_spo} shows an example of Bi$_2$Se$_3$ for the SGOs of space group (SG) 166. They are given by $Det~(\pm 1)$,  $\omega$, and $\vec n~(n_x,n_y,n_z)$ and $\bv$ (v$_1\bt_1$,v$_2\bt_2$,v$_3\bt_3$) with $\bt_1,\bt_2,\bt_3$ primitive lattice vectors. The $-1$ value of $Det$ indicates that the operator is a roto-inversion.
Actually, the listed SGOs depend on the lattice vectors. Primitive lattice vectors ($\bt_1,\bt_2,\bt_3$)  and primitive reciprocal lattice vectors~($\bg_1,\bg_2,\bg_3$) are read from OUTCAR, also shown in Fig.~\ref{fig:bise_lat} for Bi$_2$Se$_3$.
It's worth noting that to be compatible with the CRTs in the BCS, the POSCAR should be given in a standard way (see more details in Appendix~\ref{sup:3}).
The O(3) and SU(2) MPs are generated in the spin-1 (under the basis of $\{\bf x,y,z\}$) and spin-1/2 (under the basis of $\{\up,\dw\}$) spaces, respectively:
\begin{eqnarray}
& R(\omega,\vec n)=Det\cdot e^{-i \omega (\vec n \cdot \vec L)},~
L_x= \left(\begin{array}{ccc}
0 & 0 & 0 \\
0 & 0 &-i \\
0 & i & 0 \end{array}\right),
L_y= \left(\begin{array}{ccc}
0 & 0 & i \\
0 & 0 & 0 \\
-i& 0 & 0 \end{array}\right),
L_z= \left(\begin{array}{ccc}
0 &-i & 0 \\
i & 0 & 0 \\
0 & 0 & 0 \end{array}\right); \label{Eq:realR}\\
& S(\omega,\vec n)=e^{-i \omega (\vec n \cdot \vec S)},~
S_x= \frac{\sigma_x}{2}=\frac{1}{2}\left(\begin{array}{cc}
0  & 1 \\
1  & 0 \end{array}\right),
S_y= \frac{\sigma_y}{2}=\frac{1}{2}\left(\begin{array}{cc}
0 &-i \\
i & 0 \end{array}\right),
S_z= \frac{\sigma_z}{2}=\frac{1}{2}\left(\begin{array}{cc}
1 & 0 \\
0 &-1 \end{array}\right). 
\label{eq:13}
\end{eqnarray}

In 3D crystals, it is more convenient to use MPs in the lattices of~($\bt_1,\bt_2,\bt_3$) in real space and  in reciprocal lattices of~($\bg_1,\bg_2,\bg_3$) in momentum space. They are given in the following convention:
\begin{eqnarray}
&\vec v= \bt_1v_1+\bt_2v_2+\bt_3v_3=(\bt_1,\bt_2,\bt_3)
\left(\begin{array}{c}
v_1 \\
v_2 \\
v_3
\end{array}\right),~
(\bt_1,\bt_2,\bt_3) \equiv 
\left(\begin{array}{ccc}
t_{1x}&t_{2x}&t_{3x} \\
t_{1y}&t_{2y}&t_{3y} \\
t_{1z}&t_{2z}&t_{3z} \\
\end{array}\right); \label{eq:14} \\
&
\vec k= k_1\bg_1+k_2\bg_2+k_3\bg_3=(k_1,k_2,k_3)
\left(\begin{array}{c}
\bg_1 \\
\bg_2 \\
\bg_3
\end{array}\right)
,~
\left(\begin{array}{c}
\bg_1 \\
\bg_2 \\
\bg_3
\end{array}\right)\equiv  
\left(\begin{array}{ccc}
g_{1x}&g_{1y}&g_{1z} \\
g_{2x}&g_{2y}&g_{2z} \\
g_{3x}&g_{3y}&g_{3z} \\
\end{array}\right). \\
&\text{ with }\left(\begin{array}{c}
\bg_1 \\
\bg_2 \\
\bg_3
\end{array}\right) 
(\bt_1,\bt_2,\bt_3) =2\pi \mathbb I_{3\times 3} \notag
\end{eqnarray}
The rotational symmetry operators acting on the vectors are transformed as:
\begin{eqnarray}
&R\vec v=(\bt_1,\bt_2,\bt_3)  Z 
\left(\begin{array}{c}
v_1 \\
v_2 \\
v_3
\end{array}\right),~ 
R\vec k=(k_1,k_2,k_3)  Z^{-1}
\left(\begin{array}{c}
\bg_1 \\
\bg_2 \\
\bg_3
\end{array}\right);~
\label{eq:rk} \\
&R( \bt_1,\bt_2,\bt_3)=(\bt_1,\bt_2,\bt_3) Z \Rightarrow Z \equiv (\bt_1,\bt_2,\bt_3)^{-1}R( \bt_1,\bt_2,\bt_3)
\label{eq:defZ}
\end{eqnarray}
Thus, rotational MPs in the lattice vectors are $3\times 3$ integer matrices ($Z$), which are defined in Eq.~(\ref{eq:defZ}). Instead of the real $R$-matrices in Cartesian coordinates in Eq.~(\ref{Eq:realR}), the integer matrices, $Z$ and $Z^{-1}$, are actually stored and used throughout the code, which are all set in the subroutine: {\ttfamily init.f90}.

If one wants to do some sub-space-group symmetry calculations, one can modify the SGOs in OUTCAR and give the correct space group number accordingly. For example, if one only wants to know parity eigenvalues of the energy bands, one can change the list of SGOs with only two lines (\ie identity and inversion symmetry) and give space group \#2 to run \pgma.

\subsection{Little group of a certain $k$-point}
The eigen-wavefunctions at a certain $k$-point only support the IRs of the little group of $\bk$, $LG(k)$. Therefore, for any given $k$-point, the program has to determine the $\bk$-little group $LG(k)$ first.
This is done in the subroutine: {\ttfamily kgroup.f90}. The $LG(k)$ are defined by Eq. (\ref{eq:lgk}). In the program, the integer matrices $Z^{-1}$ and Eq. (\ref{eq:rk}) in momentum space are used.

\subsection{Character tables for $\bk$-little groups}
Currently, there are two conventions of CRTs for $k$-little groups.
In the first convention, the $k$ points are labeled by the IRs of the PNGs, since IRs of the space group can be expressed as IRs of the corresponding point group multiplied by a phase factor. They are suitable either for symmorphic SGs, or the inner $k$-points (not on the BZ boundary/surface) for the non-symmorphic SGs. 
The CRTs of PNGs are given in the Ref.~\cite{cornwell1984group,streitwolf1971group}, which have been implemented in the program {\ttfamily irrep} of the WIEN2k package~\cite{blaha2001wien2k,persson1999}.
In the second convention, all the CRTs for $k$-points of all 230 SGs are listed on the BCS~\cite{stokes2013tabulation}. Therefore, the program \pgma~works for all $k$-points in 230 SGs.
The CRTs are retrieved from the inputs of the BCS, which is done by the subroutine: {\ttfamily nonsymm.f90}.

\begin{figure}[htb]
  \raggedright
  \includegraphics[scale=0.34]{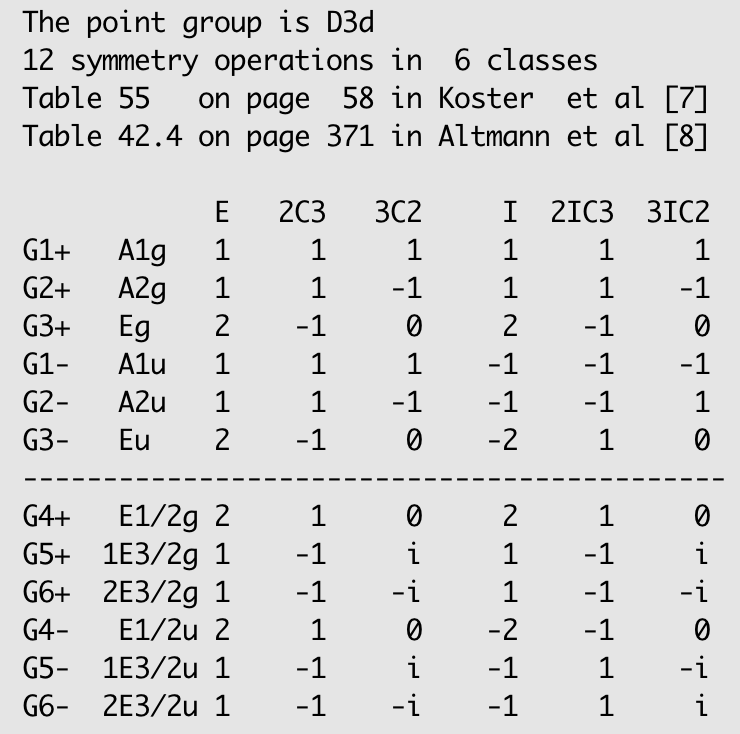}
  \caption{The character table of point group $D_{3d}$, which is used to determine the IRs (\ie PNG convention) for the energy bands at $\Gamma$ of Bi$_2$Se$_3$ in {\ttfamily irrep} of the WIEN2k package.}
  \label{fig:bise_ptg}
\end{figure}

\begin{figure}[htb]
  \raggedright
  \includegraphics[scale=0.38]{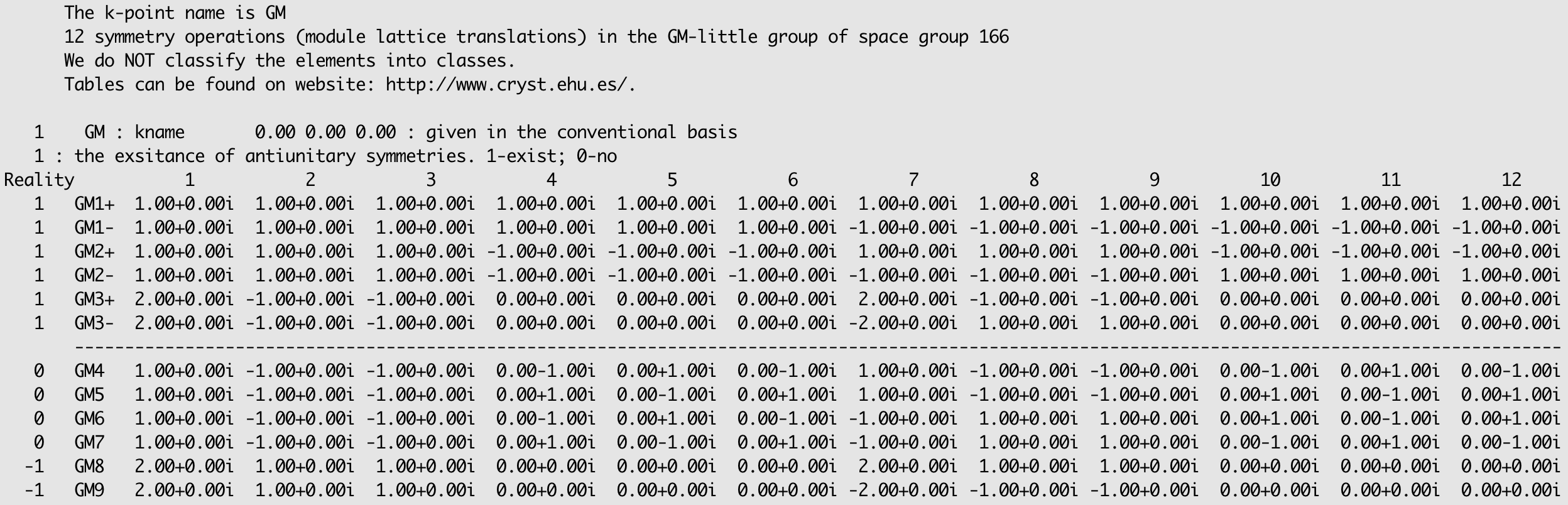}
  \caption{The character table of $\Gamma$-little group in SG 166 on the BCS, which is used to determine the IRs (\ie BCS convention) for the energy bands at $\Gamma$ of Bi$_2$Se$_3$ in the program \pgma.}
  \label{fig:bise_bcs}
\end{figure}

As an example, consider the $\Gamma$ point of Bi$_2$Se$_3$. Fig.~\ref{fig:bise_ptg} shows the CRT of the point group $D_{3d}$ in the PNG convention. Fig.~\ref{fig:bise_bcs} shows the CRT in the BCS convention. Both tables can be used to determine the IRs at $\Gamma$ in SG 166. In the table of Fig.~\ref{fig:bise_bcs}, the first and two columns show the reality and the BCS labels of IRs, respectively. The following columns indicate the characters of different SGOs. 
The reality of an IR is given by the definition~\cite{cornwell1984group,streitwolf1971group}:
\begin{eqnarray}
&\frac{1}{|\bG|}\sum_{j=1}^{|\bG|} \chi(G_j^2)
=\begin{cases}
\begin{array}{ccc}
1  &\text{potentially real   ,}&\text{ case (a)}\\
0  &\text{essentially complex,}&\text{ case (b)} \\
-1 &\text{pseudo-real        ,}&\text{ case (c)}
\end{array}\end{cases}
\end{eqnarray}
where $G_{j}$ is an element of the group $G$, and $|G|$ is the rank of the group $G$.
In a MSG, the group $G$ is defined as the unitary part of the MSG.
In case (a), the IR is equivalent to its complex representation, and also equivalent to a real representation.
In case (b), the IR is not equivalent to its complex representation.
In case (c), the IR is equivalent to its complex representation, but not to a real representation.

In the type-II MSGs, including pure time-reversal symmetry (TRS), the existence of anti-unitary SGOs in the $k$-little group is indicated at the beginning of the character table (Fig.~\ref{fig:bise_bcs}). 
In the absence of SOC (integer spin), TRS doubles the degeneracy of IRs in cases (b) and (c); while in the presence of SOC (half-integer spin), it doubles the degeneracy of the IRs in cases (a) and (b).

\vspace{-0.1in}

\vspace{-0.2in}

\subsection{Determination of irreducible representations}
After the normalization of the PW-based pseudo-wavefunctions in VASP, the traces of MPs of SGOs can be computed via Eq. (\ref{eq:pw}), which are done in the subroutine {\ttfamily chrct.f90}. By comparing the obtained traces and the characters of the CRTs, the IRs can be determined (see different versions in Appendix~\ref{sup:5}).

\section{Capabilities of \pgma}
\label{sec:4}
In the study of the properties of a material, the determination of IRs of computed electron bands is of great interest to diagnose the band crossing/anti-crossing, degeneracy and band topology. 
In the WIEN2k package, the program {\ttfamily irrep} classifies the IRs in PNG symmetries, which then excludes the possibility to describe certain BZ surface $k$-points for nonsymmorphic crystals.
Therefore, the demand to determine the IRs for all the $k$-point in all 230 SGs is still unsatisfied. With the CRTs from the BCS, the program -- {\ttfamily irvsp} -- is developed to meet this demand with the interface to the VASP package.
The associated library -- {\ttfamily irrep\_bcs.a} -- can be easily linked to by other \emph{ab-initio} packages.
The obtained IRs are labeled in the convention of the BCS notation, which can be directly compared with the elementary band representations (EBRs) of the TQC theory, to further check the topology of a set of bands in materials.

\section{Installation and usage}
\label{sec:5}
In this section, we will show how to install and use the \pgma~software package. This program is an open source free software package. It is released on Github under the GNU Lesser General Public License, \url{https://www.gnu.org/licenses/lgpl-3.0.html}, and it can be downloaded directly from the public code archive: \url{https://github.com/zjwang11/irvsp/blob/master/src\_irvsp\_v2.tar.gz}.

To build and install \pgma, only a Fortran 90 compiler is needed. 
The downloaded \pgma~software package is likely a compressed file with a {\ttfamily zip} or {\ttfamily tar.gz} suffix. One should uncompress it first, then move into the {\ttfamily src\_irvsp\_v2} folder. After setting up the Fortran compiler in the {\ttfamily Makefile} file, the executable binary \pgma~can be compiled by typing the following command in the current directory ({\ttfamily src\_irvsp\_v2}):

\lstset{language=bash, keywordstyle=\color{blue!70}, basicstyle=\ttfamily, frame=shadowbox}
\begin{lstlisting}
  $ ./configure.sh 
  $ source ~/.bashrc 
  $ make 
\end{lstlisting}
Before running \pgma, the user must provide two consistent files: WAVECAR and OUTCAR. The two files are generated by the VASP package in fixed format. 
It is designed to be simple and user friendly. After a running of VASP with WAVECAR and OUTCAR output, the program \pgma~can be run immediately.
Giving a correct space group number ($sgn~\in~[1,230]$) and a set of energy bands (from the $m$th band to the $n$th band), the program can be executed by the following command:

\lstset{language=bash, keywordstyle=\color{blue!70}, basicstyle=\ttfamily, frame=shadowbox}
\begin{lstlisting}
  $ irvsp -sg $sgn [-nb $m $n] > outir &
\end{lstlisting}

\section{Examples}
\label{sec:6}
Very recently, the codes \href{https://github.com/zjwang11/irvsp/blob/master/src\_trace\_v1.tar.gz}{\ttfamily vasp2trace} and \href{https://www.cryst.ehu.es/cryst/checktopologicalmat}{\ttfamily CheckTopologicalMat} have been designed for TQC in the Ref. \cite{vergniory2019complete}. However, they are not suitable for non-maximal HSK points. 
In fact, {\ttfamily vasp2trace} is extracted from \pgma~to interface with {\ttfamily CheckTopologicalMat}.
Here, we take topological materials PdSb$_2$ and Bi as examples to show how to study topological properties of new materials with \pgma.
The necessary files for these materials are given as the examples in the archive \href{https://github.com/zjwang11/irvsp/blob/master/src\_irvsp\_v2.tar.gz}{\ttfamily src\_irvsp\_v2.tar.gz} .

\begin{figure}[htb]
  \raggedright
  \includegraphics[scale=0.4]{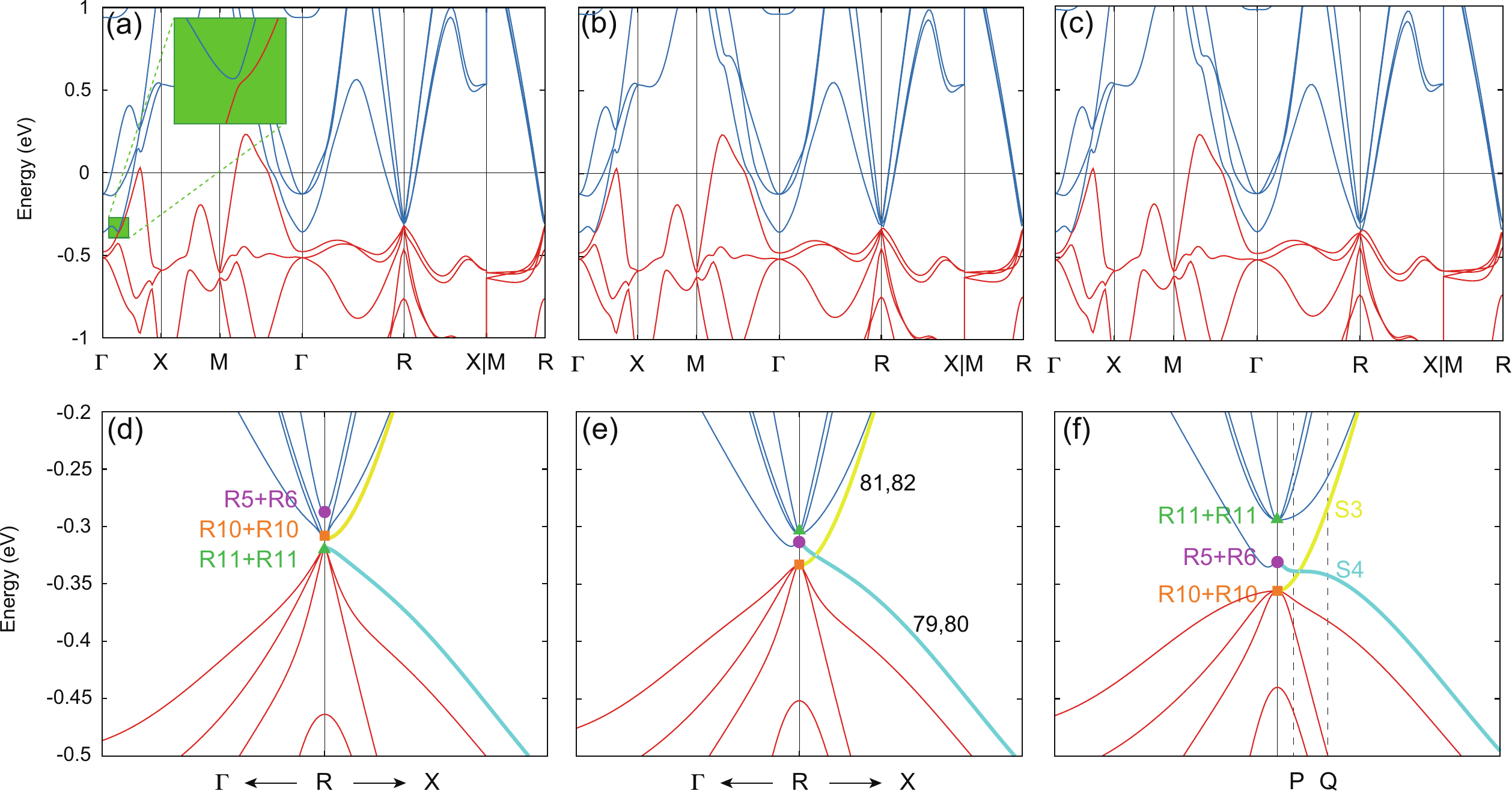}
  \caption{Electronic band structures of PdSb$_2$ without strain (a), with 0.31\% (b) and 0.62\% (c) tensile strains. Panels (d-f) are the zoom-in plots of panels (a-c) near the R point. In our calculations, the total number of electrons is 80.
  The different IRs at R are labeled by green triangles (R11+R11), orange squares (R10+R10) and purple circles (R5+R6). The two crossing bands along R--X belong to S3 (yellow) and S4 (cyan), respectively. P~$[0.4912~(\frac{2\pi}{a}),0.5000~(\frac{2\pi}{a}),0.4912~(\frac{2\pi}{a})]$ and Q $[0.4722~(\frac{2\pi}{a}),0.5000~(\frac{2\pi}{a}),0.4722~(\frac{2\pi}{a})]$ are two points near the crossing point on the R--X line.
  }
  \label{fig:PdSb2}
\end{figure}

\vspace{-0.1in}
\subsection{PdSb$_2$}
\begin{figure}[!t]
  \raggedright
 \subfigure[]{
  \includegraphics[scale=0.5]{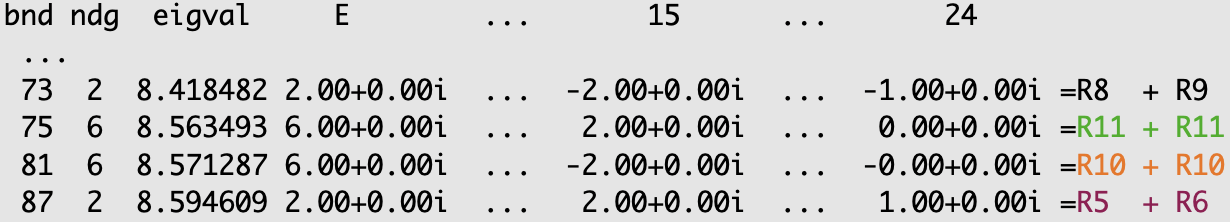}
 }
 \subfigure[]{
  \includegraphics[scale=0.5]{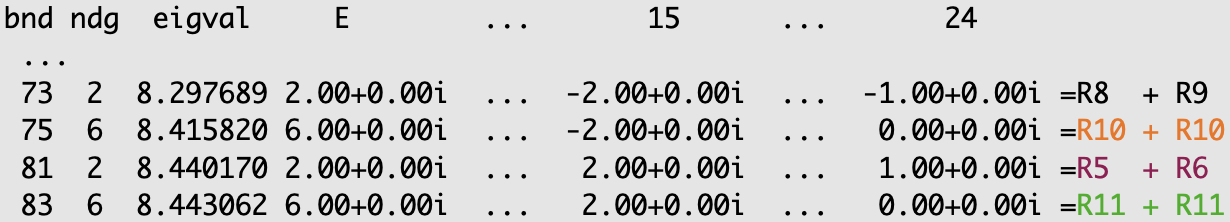}
 }
 \subfigure[]{
  \includegraphics[scale=0.5]{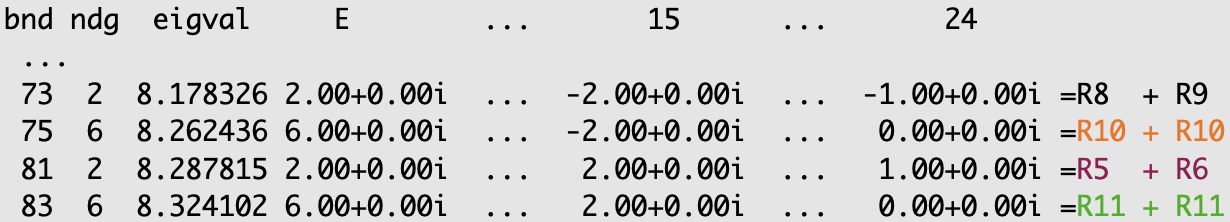}
 }
  \caption{The IRs at R are determined by the program \pgma. The CRT of the R-little group is shown in Fig.~\ref{fig:figs1} in Appendix~\ref{sup:4}. 
The first three columns stand for the band indices, degeneracies, and the energies (without subtracting the Fermi energy E$_F$). The following columns indicate the traces (characters) of the corresponding space group operators (listed as ``E, 2, $\dots$, 24"). The assigned IR labels are output to the right of the equality sign ``=".  The (a), (b), (c) panels are the obtained results for the three crystal structures, respectively.}
  \label{fig:Rrep}
\end{figure}

PdSb$_2$ was predicted to be a candidate hosting sixfold-degenerate fermions because of nonsymmorphic symmetry~\cite{bradlyn2016beyond,chapai2019fermions}. The crystal of PdSb$_2$ is a cubic structure of SG 205.
We adopt the experimental lattice constant $a$~\cite{pratt1968x,furuseth1965redetermined,brese1994bonding} and fully relax the coordinates of inner atomic positions.
In the obtained band structure (BS) along the high-symmetry lines (Fig.~\ref{fig:PdSb2}(a)), we note that there is a tiny gap ($\sim 10$ meV) between two sixfold degeneracies at R. Then, we want to know the corresponding IRs of two sixfold degeneracies and how they are going to evolve under strains. For this purpose, we performed the calculations with different tensile strains (\ie $\Delta a/a=0.31$\% and $\Delta a/a=0.62$\%). Their electronic band structures are shown in Figs.~\ref{fig:PdSb2} (b) and (c), respectively.
Comparing with the strain-free BS in Fig.~\ref{fig:PdSb2}(a), we find that the overall BS doesn't change very much, except for the R point. The zoom-in plots around R are shown in lower panels of Fig.~\ref{fig:PdSb2}.
The R point is a $k$-point with nonsymmorphic symmetry in SG 205, where IRs of the space group can not be expressed as IRs of the corresponding point group multiplied by a phase factor.
By running \pgma, the IRs at R are obtained. Figs.~\ref{fig:Rrep} (a-c) show the results of IRs for the low-energy bands. The number of total electrons is 80 for PdSb$_2$. It is shown that the energy ordering of electron bands is changed at $R$ under tiny strains.

The IRs at all the maximal HSK points can be computed directly by \pgma. The trace file -- {\ttfamily trace.txt} will be generated if only maximal HSK points are given in KPOINTS. By directly comparing these obtained IRs with the EBRs of the TQC theory (released on the BCS) and solving the compatibility relations, we can find that it is a topological insulating phase without strain, while it's a symmetry-enforced semimetallic phase with tiny tensile strains.

\begin{figure}[!t]
  \raggedright
  \subfigure[]{
  \includegraphics[scale=0.48]{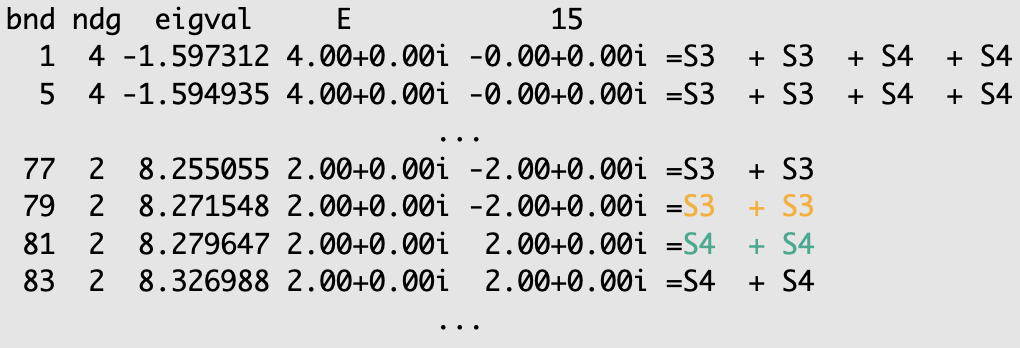}
  }
  \subfigure[]{
    \includegraphics[scale=0.48]{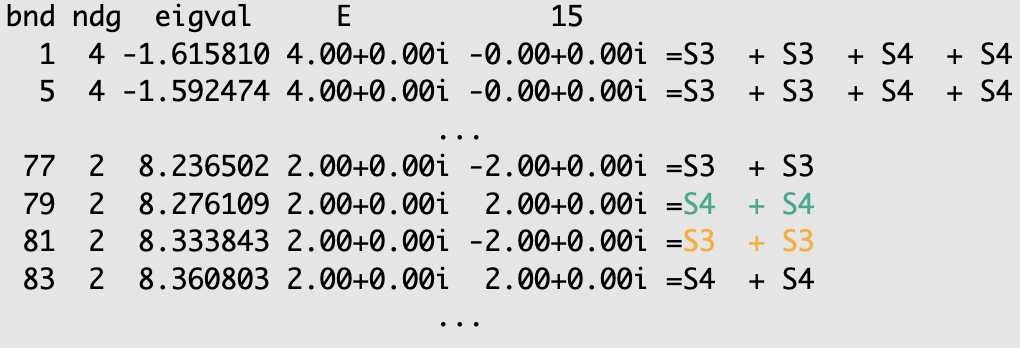}
  }
  \caption{The IRs are obtained by \pgma~for P (a) and Q (b) as depicted in Fig.~\ref{fig:PdSb2}(f).}
  \label{fig:RXrep}
\end{figure}

To further obtain the crossing points in the system, we computed the IRs along the R--X line (named S [$u,0.5,u$] in units of $\frac{2\pi}{a}$). These points are also non-symmorphic, which are on the boundary of the 3D BZ for SG 205. 
The CRT for the S point is listed in Fig.~\ref{fig:figs2}. For the P and Q points in Fig.~\ref{fig:PdSb2}(f) of the strained crystal, we show the results of obtained IRs in Fig.~\ref{fig:RXrep}. At the P point, the 79-80 degenerate bands are assigned to ``S3+S3", while 81-82 degenerate bands are assigned to ``S4+S4". However, at the Q point, the results are in the opposite way. 
Without doing further calculations with a denser kmesh between P and Q points, we can still conclude that it's a real 4-fold crossing along R--X on the BZ boundary, which is robust against SOC. The double degeneracy is due to the presence of TRS. The symmetry \#15 is the operator $g_y\equiv \{M_y|0\frac{1}{2}\frac{1}{2}\}$. Therefore, the doubly-degenerate bands have the same $g_y$ eigenvalue ($\{S3,S3\}$ or $\{S4,S4\}$), and the 4-fold crossing point along R--X is protected by $g_y$ symmetry. As a result, the crossing 4-fold points actually form a Dirac nodal ring on the BZ boundary. Considering the full symmetry of SG 205, we conclude that there are three Dirac nodal rings in PdSb$_2$ with tiny strains, which can be further checked in experiments in the future.

\begin{table}[!b]
  \captionstyle{centerlast}
  \caption{
The IRs at maximal HSK points obtained by \pgma\ for Bismuth. ``($p$)" indicates the degeneracy of the bands, while ``[$q$]" indicates the total number of the computed bands at the $k$-point.
  }
  \begin{tabular}{p{0.8cm}p{6cm}}
\hline
\hline
   HSK &  six valence bands \\
\hline
    GM& GM8 (2); GM8 (2); GM4 GM5 (2); [6] \\
    T & T9  (2); T8  (2); T6  T7  (2); [6] \\
    F & F3  F4  (2); F5  F6  (2); F5  F6  (2); [6] \\
    L & L3  L4  (2); L5  L6  (2); L5  L6  (2); [6] \\
\hline
\hline
  \end{tabular}
  \label{tab:birep}
\end{table}

\subsection{Bismuth}
\begin{figure}[!t]
  \raggedright
  \includegraphics[height=2.8cm]{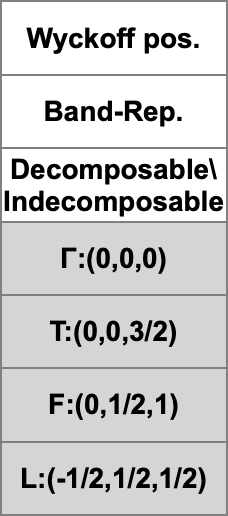}
  \includegraphics[height=2.8cm]{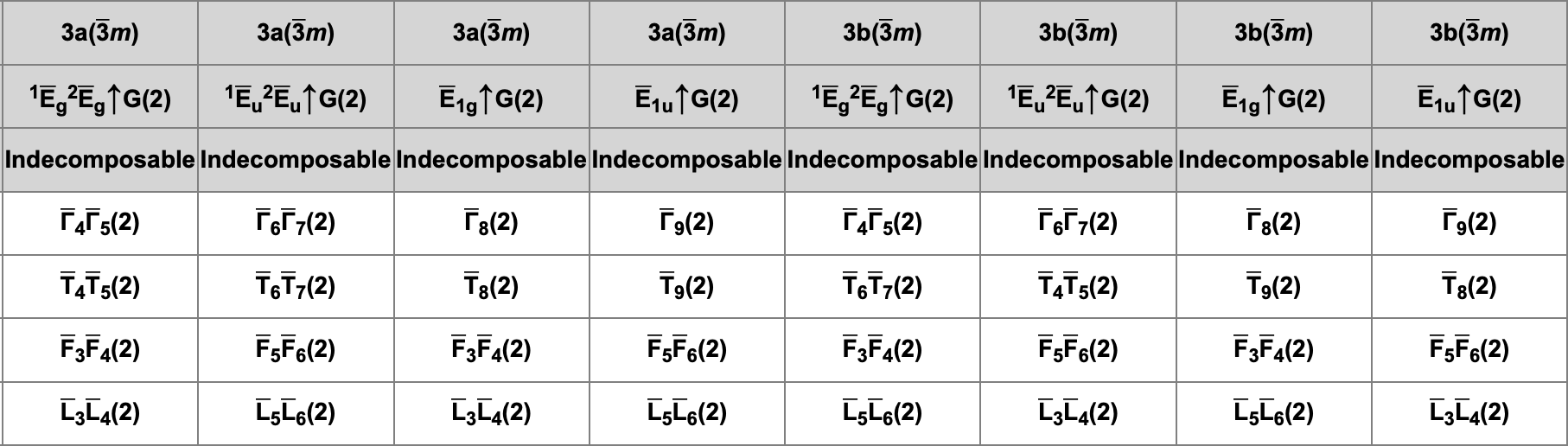}
  \includegraphics[height=2.8cm]{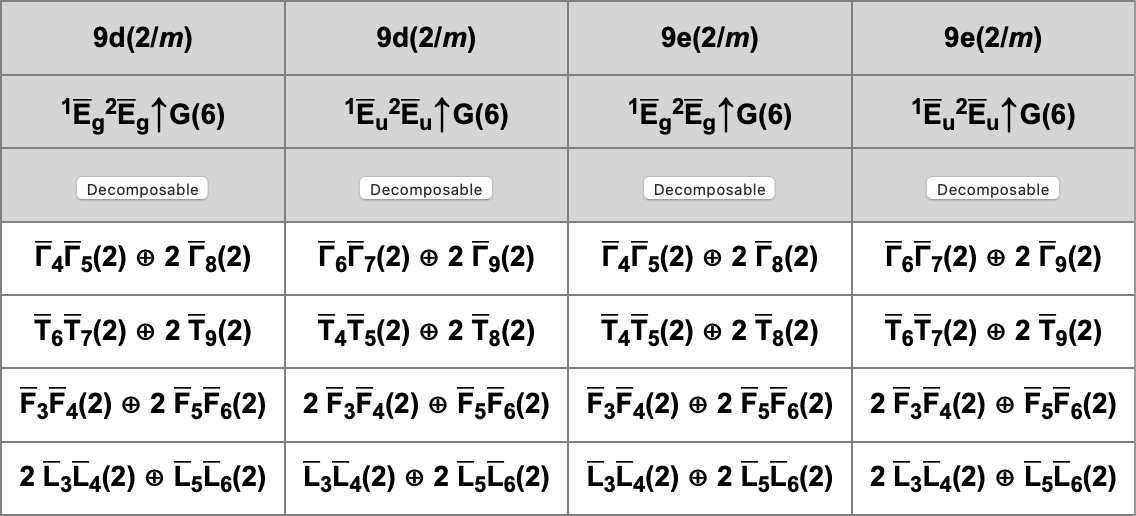}
  \caption{A complete list of the EBRs of space group 166 in the presence of SOC. Each EBR is defined by a Maximal Wyckoff site ($n$x) and an IR of its site-symmetry group, which are indicated by the first and second rows, respectively. Then, the following rows present the IRs at the Maximal HSK points.}
  \label{fig:166ebr}
\end{figure}

\vspace{0.1in}
As aforementioned, with the IRs at maximal HSK points obtained by \pgma, we can further check the topology by comparing them with the EBRs of the TQC theory. Here, we will take Bi as an example to briefly introduce the process.
The element Bismuth has the rhombohedral structure of SG 166. The maximum HSK points of SG 166 are listed on the BCS, as $\Gamma$(GM), T, F, L. After performing the \emph{ab-initio} calculations to obtain the eigen-wavefunctions at maximal HSK points, the obtained IRs of the occupied bands are given in Table~\ref{tab:birep}. From the TQC and BCS, the EBRs of SG 166 are obtained, as shown in Fig.~\ref{fig:166ebr}. As there are only six valence bands, we can find that they do not belong to any EBR induced from the $9d$ or $9e$ Wyckoff position. In the EBRs induced from the $3a$ and $3b$ Wyckoff positions, we can find that the number of the pairs of F5F6 at F has to be the same as the total number of the IRs GM9 and GM6GM7 at $\Gamma$. 
In Bismuth, the obtained IRs have three IRs of F5F6, but neither GM9 nor GM6GM7. Therefore, the occupied bands of Bisumth can not be expressed as any sum of EBRs in SG 166. In other words, it has to be topological~\cite{schindler2018higher}.

\section{Conclusions}
In summary, we present an open-source software package -- \pgma~-- that determines the IRs of electronics states in the VASP. It is very user-friendly and is written in Fortran 90/77, showing a powerful function to analyze the IRs for all the $k$-points in all 230 SGs, including nonsymmorphic crystals. The associated library -- {\ttfamily irrep\_bcs.a} -- can be interfaced with other DFT packages.
We show how to use it to identify IRs and further get the topological property for a new material. As an example, we explore a topological material PdSb$_2$, whose topology is very sensitive to the lattice parameter. Under tiny strains, it is identified as a four-fold Dirac nodal-line metal.

\ \\
\noindent \textbf{Acknowledgments}
We thank Dr.~Peter Blaha and Dr.~Luis Elcoro for sharing the character tables of 32 point groups implemented in the WIEN2k package and the character tables of 230 SGs for all $k$-points on the BCS.
This work was supported by the National Nature Science Foundation of China (Grant No. 11974395), the Strategic Priority Research Program of Chinese Academy of Sciences (Grant No. XDB33000000), the Center for Materials Genome and the CAS Pioneer Hundred Talents Program.
Q. S. W. acknowledges the support of NCCR MARVEL.


\ \\

\clearpage

\beginsupplement{}
\setcounter{section}{0}
\renewcommand{\thesubsection}{\arabic{subsection}}
\section*{APPENDIX}
\subsection{{\ttfamily tbbox.in} for Bi$_2$Se$_3$}
\label{sup:1}
{\ttfamily
\ \\
  case = soc   ! lda or soc \\
 \\
 proj: \\
 orbt = 2 \\
 ntau = 5 \\
 0.39900000  0.39900000  0.39900000 1 3  ! x1, x2, x3, itau, iorbit \\
 0.60100000  0.60100000  0.60100000 1 3 \\
 0.20600000  0.20600000  0.20600000 2 3 \\
 0.79400000  0.79400000  0.79400000 2 3 \\
 0.00000000  0.00000000  0.00000000 2 3 \\
 end projections \\
 \\
 kpoint: \\
  kmesh = 10 \\
  Nk = 4     \\ 
 0.00000000    0.00000000    0.00000000  ! k1: y1,y2,y3 \\
 0.50000000    0.50000000    0.50000000  ! k2 \\
 0.50000000    0.50000000    0.00000000  ! k3 \\
 0.00000000    0.50000000    0.00000000  ! k4 \\
 end kpoint\_path \\
 \\
 unit\_cell: \\
 \begin{tabular}{r@{.}lr@{.}lr@{.}lcr@{.}lr@{.}lr@{.}lc }
  1&194537707 & -2&069000000 & 9&546666657   &&  0&139523990 & -0&241662639 & 0&034916201 & \\
  1&194537707 & 2&069000000  & 9&546666657   &&  0&139523990 & 0&241662639 & 0&034916201 & \\
 -2&389075414 & 0&000000000  & 9&546666657   &&  -0&279047979 & 0&000000000 & 0&034916201 & \\ 
 \end{tabular}
 \begin{tabular}{rr@{.}lr@{.}lr@{.}lr@{.}lr@{.}lr@{.}lr@{.}lr@{.}l}
  1  &   1&000000  &   0&000000   &  1&000000  &   0&000000   &  0&000000  &   0&000000  &   0&000000   &  0&000000 \\
  2  &  -1&000000  &   0&000000   &  1&000000  &   0&000000   &  0&000000  &   0&000000  &   0&000000   &  0&000000 \\
  3  &   1&000000  & 180&000000   &  0&866025  &   0&500000   &  0&000000  &   0&000000  &   0&000000   &  0&000000 \\
  4  &  -1&000000  & 180&000000   &  0&866025  &   0&500000   &  0&000000  &   0&000000  &   0&000000   &  0&000000 \\
  5  &   1&000000  & 120&000000   &  0&000000  &   0&000000   & -1&000000  &   0&000000  &   0&000000   &  0&000000 \\
  6  &  -1&000000  & 120&000000   &  0&000000  &   0&000000   & -1&000000  &   0&000000  &   0&000000   &  0&000000 \\
  7  &   1&000000  & 179&999999   &  0&000000  &   1&000000   &  0&000000  &   0&000000  &   0&000000   &  0&000000 \\
  8  &  -1&000000  & 179&999999   &  0&000000  &   1&000000   &  0&000000  &   0&000000  &   0&000000   &  0&000000 \\
  9  &   1&000000  & 120&000000   &  0&000000  &   0&000000   &  1&000000  &   0&000000  &   0&000000   &  0&000000 \\ 
 10  &  -1&000000  & 120&000000   &  0&000000  &   0&000000   &  1&000000  &   0&000000  &   0&000000   &  0&000000 \\
 11  &   1&000000  & 180&000000   &  0&866025  &  -0&500000   &  0&000000  &   0&000000  &   0&000000   &  0&000000 \\
 12  &  -1&000000  & 180&000000   &  0&866025  &  -0&500000   &  0&000000  &   0&000000  &   0&000000   &  0&000000 \\ 
 \end{tabular}\\
end unit\_cell \\
}

\begin{table}[!b]
  \captionstyle{centerlast}
  \caption{
  Besides the \href{https://github.com/zjwang11/irvsp/blob/master/src\_irvsp\_v2.tar.gz}{\ttfamily src\_irvsp\_v2.tar.gz} code mainly discussed in the main text, there are more codes developed, which are available in the repository: \url{https://github.com/zjwang11/irvsp/}.  Different versions of the codes are developed based on the different types of the WFs and conventions of the CRTs. 
  }
  \label{tab:src}
  \begin{tabular}{|c|c|c|}
  \hline
\diagbox{WFs}{CRTs} & PNG  & BCS \\
  \hline
  PW basis & \href{https://github.com/zjwang11/irvsp/blob/master/src\_irvsp\_v1.tar.gz}{\ttfamily src\_irvsp\_v1.tar.gz} &  \href{https://github.com/zjwang11/irvsp/blob/master/src\_irvsp\_v2.tar.gz}{\ttfamily src\_irvsp\_v2.tar.gz} \\
  \hline
  TB  basis & \href{https://github.com/zjwang11/irvsp/blob/master/src\_ir2tb\_v1.tar.gz}{\ttfamily src\_ir2tb\_v1.tar.gz} &  \href{https://github.com/zjwang11/irvsp/blob/master/src\_ir2tb\_v2.tar.gz}{\ttfamily src\_ir2tb\_v2.tar.gz} \\
  \hline
  \end{tabular}
\end{table}

\newpage
\subsection{The brief description of \pgmb}
\label{sup:2}
Based on the different types of the WFs and conventions of the CRTs, different versions of the codes are developed, as shown in Table \ref{tab:src}. The program \href{https://github.com/zjwang11/irvsp/blob/master/src\_ir2tb\_v1.tar.gz}{\ttfamily ir2tb} is based on the TB WFs. BLAS and LAPACK linear algebra libraries are needed to diagonalize the TB Hamiltonian.
To compile \pgmb, one needs to copy the library {\ttfamily irrep\_bcs.a} to the folder {\ttfamily src\_ir2tb\_v2} and type the following command:
\begin{lstlisting}
  $ make 
\end{lstlisting}

The program \pgmb\ needs two input files: {\ttfamily tbbox.in} and {\ttfamily case\_hr.dat}.
The {\ttfamily case\_hr.dat} file, containing the TB parameters in Wannier90 format~\cite{wannier90rmp}, may be generated by the software Wannier90~\cite{mostofi2014updated} with symmetrization~\cite{gresch2018automated}, or generated by users with a toy TB model, or generated from Slater-Koster method~\cite{sk1954} or a discretization of $k\cdot p$ model onto a lattice~\cite{kpmethod}. 

The {\ttfamily tbbox.in} file provides detailed information about the TB Hamiltonian (\ie the {\ttfamily case\_hr.dat} file), which is summarized in Table~\ref{tab:tb}. It is an essential input for the program \pgmb.  
The tag {\ttfamily case = lda} ({\ttfamily case = soc}) indicates that the TB Hamiltonian does not (does) have the SOC effect. The {\ttfamily lda/soc\_hr.dat} is needed accordingly.
In the {\ttfamily proj} block, {\ttfamily orbt = 1 or 2} indicates the convention of the local orbital ordering on each atom. The local orbitals in convention 1 are listed in Table~\ref{tab:conv1}, while those in convention 2 are in the order as implemented in Wannier90. {\ttfamily ntau} indicates the total number of the atoms in the TB Hamiltonian, which also means how many lines follow in this block.
The local orbitals of the TB Hamiltonian are provided by : {\ttfamily x1,x2,x3, itau, iorbit}. {\ttfamily x1,x2,x3} stand for the positions of atoms: $\tau_i=(x_1\bt_1,x_2\bt_2,x_3\bt_3)$; {\ttfamily itau} stand for the kinds of atoms; and {\ttfamily iorbit} stand for the total numbers of local orbitals on different atoms. So far, {\ttfamily iorbit} is limited to the values of \{1,3,5,4,6,7,8,9\}, whose detailed orbital informations are provided in Table~\ref{tab:conv1}. 
In the case of {\ttfamily case = soc}, the local orbitals will be doubled automatically: the first half are spin-up and the second half are spin-down.
In the {\ttfamily kpoint} block, the $k$-path is given as $k_1$ -- $k_2$ -- $\dots$ -- $k_N$ with {\ttfamily kmesh} points on each segment.
The {\ttfamily unit\_cell} block gives the lattice vectors and reciprocal lattice vectors in first three lines, followed by space group operators, which are the same lines as \pgma~reads in OUTCAR file.

\begin{table}[!h]
  \captionstyle{centerlast}
  \caption{
A brief summary of {\ttfamily tbbox.in}.
  }
  \begin{tabular}{p{5.1cm}|p{9cm}p{0.2cm}c}
  \hline
  \hline
  Comments & Descriptions \\
  \hline
  ! lda or soc &  lda: nspin=1 (without SOC); soc: nspin=2 (with SOC) \\
  ! x1,x2,x3,itau,iorbit & defining $\tau_i=(x_1\bt_1,x_2\bt_2,x_3\bt_3)$, iorbit $\in~\{1,3,4,5,6,7,8,9\}$\\ 
  ! k1: y1,y2,y3 & defining $\bk_1=(y_1\bg_1,y_2\bg_2,y_3\bg_3)$; kpath is along $\bk_1-\bk_2-\dots-\bk_N$. \\
  ! b1x b1y b1z; g1x g1y g1z & defining $\bt_1=(b_{1x}\hat x,b_{1y}\hat y,b_{1z}\hat z)$; $\bg_1=2\pi(g_{1x}\hat x,g_{1y}\hat y,g_{1z}\hat z)$\\
  ! SN,Det,omega,nx,ny,nz,v1,v2,v3&defining ${\cal O}_s=\{R_s|\bv_s\}$ with $R_s=Det\cdot e^{-i \omega (\vec n \cdot \vec L)}$ and $\bv_s=(v_1\bt_1,v_2\bt_2,v_3\bt_3)$. SN stands for the sequential number.\\
  \hline
  \hline
  \end{tabular}
  \label{tab:tb}
\end{table}

\begin{table}[!h]
  \captionstyle{centerlast}
  \caption{
  The local orbitals in convention 1 (\ie {\ttfamily orbt = 1}) are given below.
  The vector $\vec L$ is given in Eq.~(\ref{Eq:realR}) in the main text, while the vectors $\vec P$ and $\vec F$ are given in Eqs.~(\ref{eq:20}--\ref{eq:23}) in Appendix~\ref{sup:2}.
  }
  \begin{tabular}{c|p{9cm}p{0.2cm}p{4cm}}
  \hline
  \hline
 iorbit & local orbitals & & $D$-matrices in Eq. (\ref{eq:Drep})  \\
  \hline
  1  &  $s$  && $D^1=1$ \\
  3  & $p_x,p_y,p_z$ &&$D^3=Det\cdot e^{-i \omega (\vec n \cdot \vec L)}$ \\
  5  & $d_{xy},d_{yz},d_{zx},d_{x^2-y^2},d_{3z^2-r^2}$ &&$D^5=e^{-i \omega (\vec n \cdot \vec P)}$ \\
  4  &$s,p_x,p_y,p_z$ && $D^4=D^1\oplus D^3$\\  
  6  &$s,d_{xy},d_{yz},d_{zx},d_{x^2-y^2},d_{3z^2-r^2}$  &&$D^6=D^1\oplus D^5$ \\ 
  8  &$p_x,p_y,p_z,d_{xy},d_{yz},d_{zx},d_{x^2-y^2},d_{3z^2-r^2}$  && $D^8=D^3\oplus D^5$ \\ 
  9  &$s,p_x,p_y,p_z,d_{xy},d_{yz},d_{zx},d_{x^2-y^2},d_{3z^2-r^2}$  && $D^9=D^1\oplus D^3\oplus D^5$ \\ 
  7  &$f_{xyz},f_{5x^3-xr^2},f_{5y^3-yr^2},f_{5z^3-zr^2},f_{x(y^2-z^2)},f_{y(z^2-r^2)},f_{z(x^2-y^2)}$ && $D^7=Det\cdot e^{-i \omega (\vec n \cdot \vec F)}$\\
  \hline
  \hline
  \end{tabular}
  \label{tab:conv1}
\end{table}

\begin{equation}
P_x=
\left(
\begin{array}{ccccc}
 0 & 0 & -i & 0 & 0 \\
 0 & 0 & 0 & -i & -i \sqrt{3} \\
 i & 0 & 0 & 0 & 0 \\
 0 & i & 0 & 0 & 0 \\
 0 & i \sqrt{3} & 0 & 0 & 0 \\
\end{array}
\right);
P_y=
\left(
\begin{array}{ccccc}
 0 & i & 0 & 0 & 0 \\
 -i & 0 & 0 & 0 & 0 \\
 0 & 0 & 0 & -i & i \sqrt{3} \\
 0 & 0 & i & 0 & 0 \\
 0 & 0 & -i \sqrt{3} & 0 & 0 \\
\end{array}
\right);
P_z=
\left(
\begin{array}{ccccc}
 0 & 0 & 0 & 2 i & 0 \\
 0 & 0 & i & 0 & 0 \\
 0 & -i & 0 & 0 & 0 \\
 -2 i & 0 & 0 & 0 & 0 \\
 0 & 0 & 0 & 0 & 0 \\
\end{array}
\label{eq:20}
\right)
\end{equation}

\begin{equation}
F_x=
\left(
\begin{array}{ccccccc}
 0 & 0 & 0 & 0 & 2i & 0 & 0 \\
 0 & 0 & 0 & 0 & 0 & 0 & 0 \\
 0 & 0 & 0 & \frac{3i}{2} & 0 & 0 & \frac{i\sqrt{15}}{2} \\
 0 & 0 & -\frac{3i}{2} & 0 & 0 & \frac{i\sqrt{15}}{2} & 0 \\
 -2i & 0 & 0 & 0 & 0 & 0 & 0\\
 0 & 0 & 0 & -\frac{i\sqrt{15}}{2} & 0 & 0 & -\frac{i}{2}\\
 0 & 0 & -\frac{i\sqrt{15}}{2} & 0 & 0 & \frac{i}{2} & 0 \\
\end{array}
\right)
\end{equation}

\begin{equation}
F_y=
\left(
\begin{array}{ccccccc}
 0 & 0 & 0 & 0 & 0 & 2i & 0 \\
 0 & 0 & 0 & -\frac{3i}{2} & 0 & 0 & \frac{i\sqrt{15}}{2} \\
 0 & 0 & 0 & 0 & 0 & 0 & 0 \\
 0 & \frac{3i}{2} & 0 & 0 & \frac{i\sqrt{15}}{2} & 0 & 0 \\
 0 & 0 & 0 & -\frac{i\sqrt{15}}{2} & 0 & 0 & \frac{i}{2}\\
 -2i & 0 & 0 & 0 & 0 & 0 & 0 \\
 0 & -\frac{i\sqrt{15}}{2} & 0 & 0 & -\frac{i}{2} & 0 & 0 \\
\end{array}
\right)
\end{equation}

\begin{equation}
F_z=
\left(
\begin{array}{ccccccc}
 0 & 0 & 0 & 0 & 0 & 0 & 2i \\
 0 & 0 & \frac{3i}{2} & 0 & 0 & \frac{i\sqrt{15}}{2} & 0 \\
 0 & -\frac{3i}{2} & 0 & 0 & \frac{i\sqrt{15}}{2} & 0 & 0 \\
 0 & 0 & 0 & 0 & 0 & 0 & 0 \\
 0 & 0 & -\frac{i\sqrt{15}}{2} & 0 & 0 & -\frac{i}{2} & 0\\
 0 & -\frac{i\sqrt{15}}{2} & 0 & 0 & \frac{i}{2} & 0 & 0 \\
 -2i & 0 & 0 & 0 & 0 & 0 & 0 \\
\end{array}
\right)
\label{eq:23}
\end{equation}

\subsection{The standard settings for POSCAR and maximal HSK points}
\label{sup:3}
The standard (default) settings of POSCAR are listed as follows:
\begin{itemize}
    \item[a)] unique axis b (cell choice 1) for SGs within the monoclinic system.
    \item[b)] obverse triple hexagonal unit cell for R SGs.
    \item[c)] the origin choice two - inversion center at (0,0,0) - for the centrosymmetric SGs.
\end{itemize}
\indent Before one is actually doing the VASP calculations, we strongly suggest that one could run the \href{https://phonopy.github.io/phonopy/}{\ttfamily phonopy} program to get the space group number and standardise the POSCAR with the following command: \\

\begin{lstlisting}
  $ phonopy --tolerance 0.01 --symmetry -c POSCAR
  $ cp PPOSCAR POSCAR
\end{lstlisting}
\indent The maximal HSK points from the BCS are given in the conventional reciprocal lattice vectors, while the lattice vectors in VASP usually are given in the primitive cell. The transformation depends on the type of the lattice.
There are only seven different types of lattices, {\em i.e.} $P,C,B,A,R,F$ and $I$. In the $X$ type, the primitive lattices ($\vec p_1$, $\vec p_2$, $\vec p_3$) are defined by a transformation matrix $M_X$.
\begin{eqnarray}
\left ( \begin{array}{ccc} \vec p_1 & \vec p_2 & \vec p_3 \end{array} \right )& = & \left ( \begin{array}{ccc} \vec c_1 & \vec c_2 & \vec c_3 \end{array} \right )  \cdot M_X
\end{eqnarray}
where $\vec c_1$, $\vec c_2$ and $\vec c_3$ are the standard conventional lattices. In the program, all the matrices $M_X$ are given as follows: 

\[
M_P =  \left ( \begin{array}{ccc} 1 & 0 & 0 \\ 0 & 1 & 0 \\ 0 & 0 & 1 \end{array} \right );~ 
M_C =  \left ( \begin{array}{ccc} 0.5 & 0.5 & 0 \\-0.5 & 0.5 & 0 \\ 0 & 0 & 1 \end{array} \right );~ 
M_B =  \left ( \begin{array}{ccc} 0.5 & 0 & -0.5 \\ 0 & 1 & 0 \\ 0.5 & 0 & 0.5 \end{array} \right );~ 
M_A =  \left ( \begin{array}{ccc} 1 & 0 & 0 \\ 0 & 0.5 & -0.5\\ 0 & 0.5 & 0.5 \end{array} \right );~
\]
\[
M_R =  \left ( \begin{array}{ccc} 2/3 & -1/3 & -1/3 \\ 1/3 & 1/3 & -2/3\\ 1/3 & 1/3 & 1/3 \end{array} \right );~ 
M_F =  \left ( \begin{array}{ccc} 0 & 0.5 & 0.5 \\ 0.5 &  0 & 0.5 \\ 0.5 & 0.5 & 0 \end{array} \right );~ \\
M_I =  \left ( \begin{array}{ccc} -0.5 & 0.5 & 0.5 \\ 0.5 & -0.5 & 0.5 \\ 0.5 & 0.5 & -0.5 \end{array} \right ).
\]

\subsection{The character tables for R-little group and S-little group}
\label{sup:4}
\begin{figure}[!htb]
  \centering 
  \subfigure{
    \includegraphics[scale=0.4]{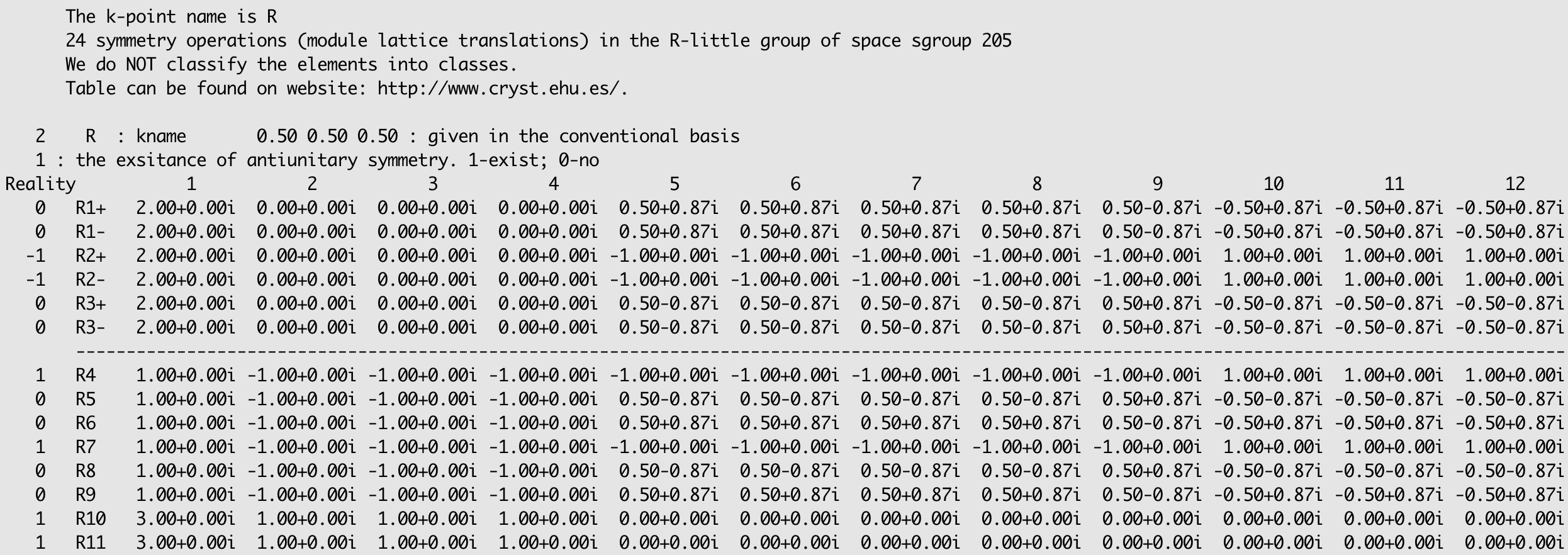}
  }
 \raggedleft
  \subfigure{
    \includegraphics[scale=0.4]{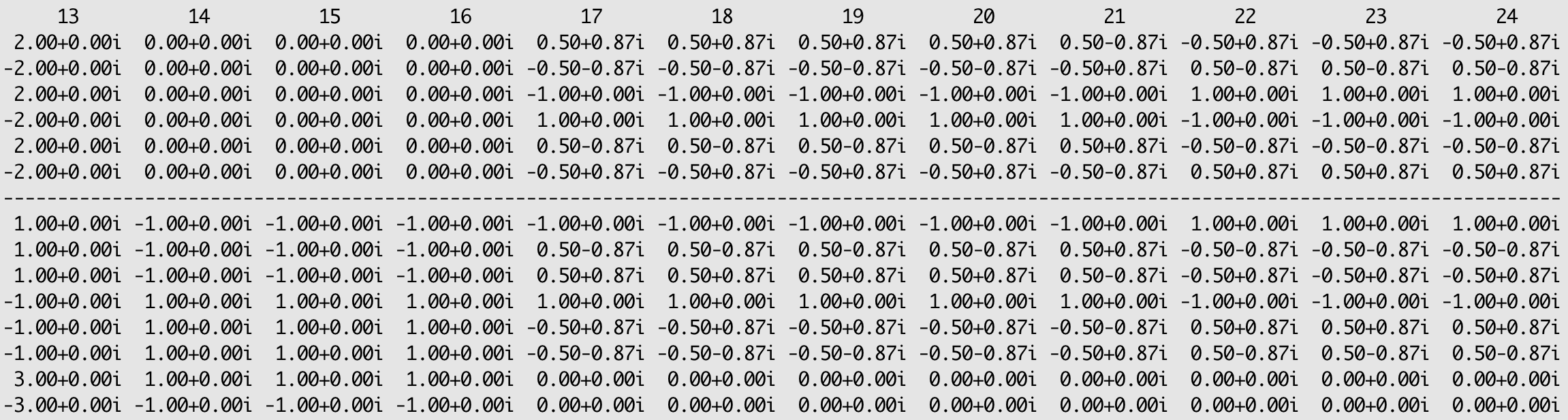}
  }
  \caption{The CRT of R-little group in the BCS convention.}
  \label{fig:figs1}
\end{figure}

\begin{figure}[!htb]
  \raggedright
  \includegraphics[scale=0.5]{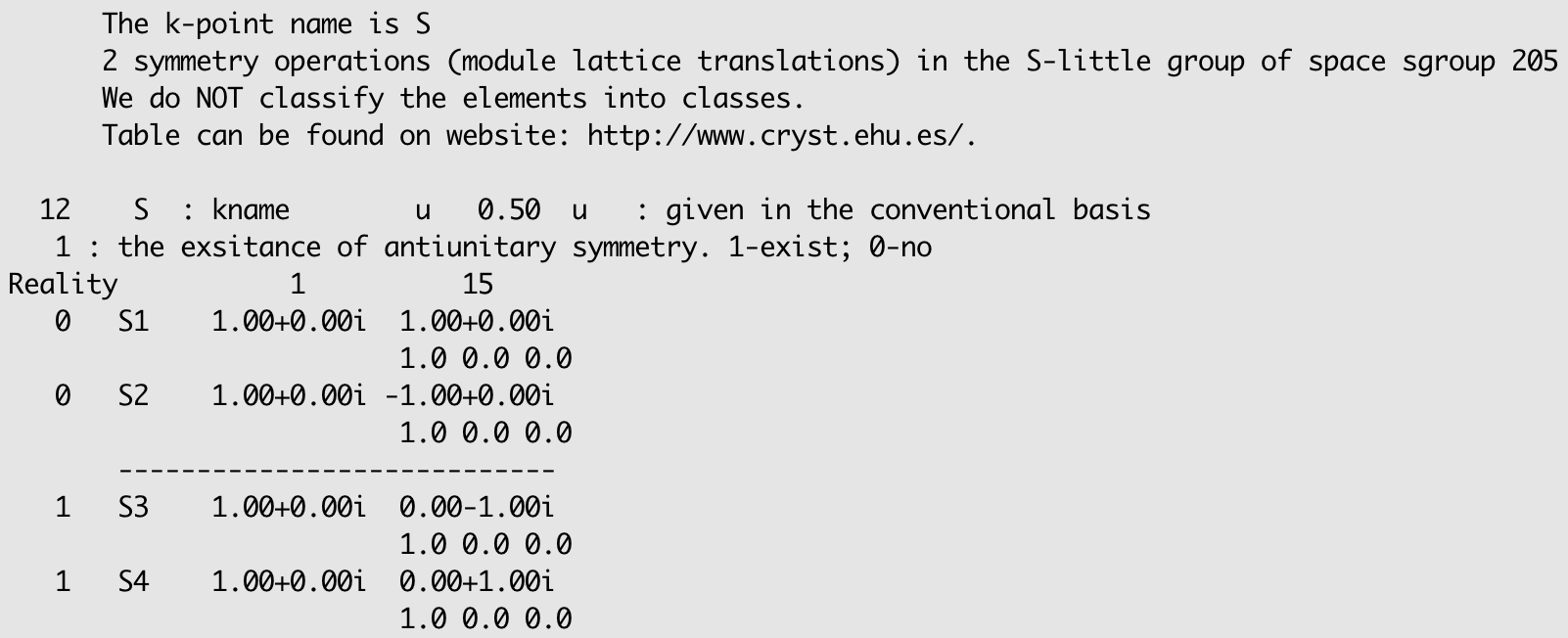}
  \caption{The CRT of S-little group in the BCS convention.}
  \label{fig:figs2}
\end{figure}

Figs.~\ref{fig:figs1} and~\ref{fig:figs2} show the character tables for R-little group and S-little group, respectively. At the $k$-point [$(u,v,w)$ given in the conventional reciprocal basis], the block {\colorbox{gray!20}{$\begin{array}{c} x+iy \\ a~b~c  \end{array}$}} in Fig.~\ref{fig:figs2} corresponds to a complex value of $(x+iy)\cdot exp[i\pi(au+bv+cw)]$.

\subsection{Other versions of \pgma}
\label{sup:5}
Four versions of \pgma~are implemented, as shown in Table~\ref{tab:ver}. Version I works similarly as {\ttfamily irrep} in the WIEN2k package and presents the IRs with PNG symmetries. This version can thus not classify the special $k$-points on the boundary of the Brillouin zone of nonsymmorphic crystals, that is, when $exp[-ik(R_s \bv_t  + \bv_s)] \neq 1$ for some $O_s$ and $O_t$ in LG(k). Version II works for those $k$-points for nonsymmorphic SGs, where version I doesn't work. Version III combines version I and II. In the (default) version IV, it works for all the $k$-points and all 230 SGs in the convention of the BCS notaion. 
One can use an optional flag {\ttfamily -v} to execute other versions of \pgma.

\lstset{language=bash, keywordstyle=\color{blue!70}, basicstyle=\ttfamily, frame=shadowbox}
\begin{lstlisting}
  $ irvsp -sg $sgn [-v $nv] [-nb $m $n] > outir &
\end{lstlisting}

\begin{table}[!h]
  \captionstyle{centerlast}
  \caption{
Four versions of \pgma~are implemented. The first column indicates the version number, the second column shows the convention of reference CRTs, and the brief description is followed in the last column.
  }
  \begin{tabular}{p{1.8cm}cp{0.1cm}p{8cm}p{0.1cm}c}
  \hline
  \hline
 Version & CRTs &&Brief description & & \\
  \hline
 version I& PNG &&It resembles an analogue of the program {\ttfamily irrep} in the WIEN2k package. && \\
  \hline
 version II&BCS &&It works only for the $k$-points, where version I does not work. && \\
  \hline
 version III& PNG\&BCS && It combines version I and version II. & & \\
  \hline
 version IV$~~~~~$ (default)& BCS&& It works for all the $k$-points and all 230 SGs, including \emph{nonsymmorphic} SGs. All the IRs are labeled in the convention of the BCS notation.
& & \\
  \hline
  \hline
  \end{tabular}
  \label{tab:ver}
\end{table}

\subsection{The library -- {\ttfamily irrep\_bcs.a} }
\label{sup:6}
The library {\ttfamily irrep\_bcs.a} is developed to be interfaced with other DFT packages. Calling the main subroutine {\ttfamily irrep\_bcs} can output the IRs labeled in the convention of BCS notation. 
The program \pgmb~is an example of calling the library mode. In other words, \pgmb\ has to be compiled by linking to the library {\ttfamily irrep\_bcs.a}. The source files of the library {\ttfamily irrep\_bcs.a} are released in the public archive: \url{https://github.com/zjwang11/irvsp/blob/master/lib\_irrep\_bcs.tar.gz}

To compile the library, one should first uncompress the archive {\ttfamily lib\_irrep\_bcs.tar.gz}, then move into the folder {\ttfamily lib\_irrep\_bcs} and type the following command:

\lstset{language=bash, keywordstyle=\color{blue!70}, basicstyle=\ttfamily, frame=shadowbox}
\begin{lstlisting}
  $ ./configure.sh
  $ source ~/.bashrc
  $ make lib 
\end{lstlisting}

The first two commands add an environment variable {\ttfamily IRVSPDATA} and the third command create the library {\ttfamily irrep\_bcs.a} in the current folder.
There are three main subroutines: {\ttfamily  irrep\_bcs, pw\_setup, tb\_setup} in the library.
Their headers  and detailed descriptions are given below ({\ttfamily dp = 8}).

\lstset{language=Fortran, keywordstyle=\color{blue!70}, basicstyle=\ttfamily, frame=shadowbox}
\begin{lstlisting}
  subroutine irrep_bcs(sgn, num_sym, &
                       rot_input, tau_input, SO3_input, SU2_input, &
                       KKK, WK, kphase, &
                       num_bands, m, n, ene_bands, &
                       spinor, dim_basis, num_basis, &
                       coeffa, coeffb, &
                       Gphase_pw, rot_vec_pw, rot_mat_tb)
\end{lstlisting}

\begin{itemize}
  \item {\ttfamily integer, intent(in) :: sgn}                                             \\ The space group number.
  \item {\ttfamily integer, intent(in) :: num\_sym}                                        \\ The number of space-group operations $\mathcal{O}_s \equiv \{R_s|{\bf v}_s\}$ (module the integer lattice translations).
  \item {\ttfamily integer, dimension(3,3,num\_sym), intent(in) :: rot\_input}             \\ The rotation part $R_s$ of space-group operations $\mathcal{O}_s$ with respect to primitive lattice vectors [\ie the matrix $Z$ in Eq.~(\ref{eq:defZ})].
  \item {\ttfamily real(dp), dimension(3,num\_sym), intent(in) :: tau\_input}              \\ The translation part ${\bf v}_s$ of space-group operations $\mathcal{O}_s$ with respect to primitive lattice vectors.
  \item {\ttfamily real(dp), dimension(3,3,num\_sym), intent(in) :: SO3\_input}            \\ The $R_s$ given in Cartesian coordinates [\ie $R(\omega, \vec{n})$ in Eq.~(\ref{Eq:realR})].
  \item {\ttfamily complex(dp), dimension(2,2,num\_sym), intent(in) :: SU2\_input}         \\ The $R_s$ given in spin-1/2 space [\ie $S(\omega, \vec{n})$ in Eq.~(\ref{eq:13})].
  \item {\ttfamily integer, intent(in) :: KKK}                                             \\ The sequential number of the given k-point.
  \item {\ttfamily real(dp), dimension(3), intent(in) :: WK}                               \\ The coordinates of the k-point with respect to primitive reciprocal lattice vectors.
  \item {\ttfamily complex(dp), dimension(num\_sym), intent(in) :: kphase}                 \\ The k-dependent phase factors due to the translation part ${\bf v}_s$ [\ie $e^{-i{\bf k}\cdot {\bf v}_s}$ in Eq.~(\ref{eq:pw}) or $e^{-i(R_s {\bf k}\cdot {\bf v}_s)}$ in Eq.~(\ref{eq:9})].
  \item {\ttfamily integer, intent(in) :: num\_bands}                                      \\ The total number of bands.
  \item {\ttfamily integer, intent(in) :: m, n}                                            \\ The IRs of the set of bands [{\it m}, {\it n}]  are computed.
  \item {\ttfamily real(dp), dimension(num\_bands), intent(in) :: ene\_band}               \\ The energy of the bands at the k-point.
  \item {\ttfamily logical, intent(in) :: spinor}                                          \\ Set to {\ttfamily .true.} if underlying electronic structure calculation has been performed with spinor wavefunctions.
  \item {\ttfamily integer, intent(in) :: dim\_basis}                                      \\ 
   The reserved number of the PW/TB basis.\\
   If {\ttfamily rot\_vec\_pw} is given, {\ttfamily dim\_basis >= num\_basis} for any k-point.\\
   If {\ttfamily rot\_mat\_tb} is given, one should set {\ttfamily dim\_basis = num\_basis}.
  \item {\ttfamily integer, intent(in) :: num\_basis}                                      \\ The number of PW or orthogonal TB basis for the given k-point (note: the number of PWs for different k-points are usually different).
  \item {\ttfamily complex(dp), dimension(dim\_basis,num\_bands),intent(in) :: coeffa}     \\ The coefficients of spin-up part of wave functions at the given k-point (note: only {\ttfamily coeffup\_basis(1:num\_basis, 1:num\_bands)} is nonzero). 
  \item {\ttfamily complex(dp), dimension(dim\_basis,num\_bands),intent(in) :: coeffb}     \\ The coefficients of spin-down part of wave functions at the given k-point if spinor is {\ttfamily .true.} (note: only {\ttfamily coeffdn\_basis(1:num\_basis, 1:num\_bands)} is nonzero). 
  \item {\ttfamily complex(dp), dimension(dim\_basis,num\_bands),intent(in), optional :: Gphase\_pw}  \\ The phase factors dependent on the PW vectors [\ie $e^{-i{\bf G}_{j^\prime}\cdot {\bf v}_s}$ in Eq.~(\ref{eq:pw})].
  \item {\ttfamily integer(dp), dimension(dim\_basis,num\_bands),intent(in), optional :: rot\_vec\_pw}  \\ The transformation vectors of space-group operations $\mathcal{O}_s$, which send the $j$th PW to the $j^\prime$th PW [\ie ${\bf G}_{j^\prime} \equiv R_s({\bf k}+{\bf G}_j)-{\bf k}$ in Eq.~(\ref{eq:pw})].
  \item {\ttfamily integer(dp), dimension(dim\_basis,dim\_basis,num\_bands),intent(in), optional :: rot\_mat\_tb}  \\ The transformation matrices of space-group operations $\mathcal{O}_s$ in the orthogonal TB basis [\ie $\overline{V(R_s{\bf k}-{\bf k})D}$ in Eq.~(\ref{eq:9})].
\end{itemize}

\lstset{language=Fortran, keywordstyle=\color{blue!70}, basicstyle=\ttfamily, frame=shadowbox}
\begin{lstlisting}
  subroutine pw_setup(WK, lattice, &
                      num_sym, det, angle, axis, tau, &
                      dim_basis, num_basis, Gvec, &
                      rot, SO3, SU2, &
                      kphase, Gphase_pw, rot_vec_pw)
\end{lstlisting}
\begin{itemize}
  \item {\ttfamily real(dp), dimension(3), intent(in) :: WK}                               \\ The coordinates of the k-point with respect to primitive reciprocal lattice vectors.
  \item {\ttfamily real(dp), dimension(3,3), intent(in) :: lattice(3,3)}                   \\ The primitive lattice vectors in Cartesian coordinates [\ie $({\bf t_1},{\bf t_2},{\bf t_3})$ in Eq.~(\ref{eq:14})].
  \item {\ttfamily integer, intent(in) :: num\_sym}                                        \\ The number of space-group operations $\mathcal{O}_s \equiv \{R_s|{\bf v}_s\}$ (module the integer lattice translations).
  \item {\ttfamily real(dp), dimension(num\_sym), intent(in) :: det}                       \\ The determination of the rotation part $R_s$ of space-group operations $\mathcal{O}_s$ [\ie $Det$ in Eq.~(\ref{Eq:realR})].
  \item {\ttfamily real(dp), dimension(num\_sym), intent(in) :: angle}                     \\ The rotation angle of space-group operations $\mathcal{O}_s$ [\ie $\omega$ in Eq.~(\ref{Eq:realR})].
  \item {\ttfamily real(dp), dimension(3,num\_sym), intent(in) :: axis}                    \\ The rotation axis of space-group operations $\mathcal{O}_s$ in Cartesain coordinates[\ie $\vec{n}$ in Eq.~(\ref{Eq:realR})].
  \item {\ttfamily real(dp), dimension(3,num\_sym), intent(in) :: tau}                     \\ The translation part ${\bf v}_s$ of space-group operations $\mathcal{O}_s$ with respect to primitive lattice vectors.
  \item {\ttfamily integer, intent(in) :: dim\_basis}                                      \\ 
   The reserved number of the PW basis ({\ttfamily dim\_basis >= num\_basis}).
  \item {\ttfamily integer, intent(in) :: num\_basis}                                      \\ The number of the PWs for the given k-point (note: {\ttfamily num\_basis} for different k-points are usually different).
  \item {\ttfamily integer, dimension(3, dim\_basis), intent(in) :: Gvec}                  \\ The plane-wave G-vector with respect to reciprocal lattice vectors [\ie $\bG_{j}$ in Eq.~(\ref{eq:3})].
  \item {\ttfamily integer, dimension(3,3,num\_sym), intent(out) :: rot}                   \\ The rotation part $R_s$ of $\mathcal{O}_s$ with respect to primitive lattice vectors [\ie the matrix $Z$ in Eq.~(\ref{eq:defZ})].
  \item {\ttfamily real(dp), dimension(3,3,num\_sym), intent(out) :: SO3}                  \\ The $R_s$ given in Cartesian coordinates [\ie $R(\omega, \vec{n})$ in Eq.~(\ref{Eq:realR})].
  \item {\ttfamily complex(dp), dimension(2,2,num\_sym), intent(out) :: SU2}               \\ The $R_s$ given in spin-1/2 space [\ie $S(\omega, \vec{n})$ in Eq.~(\ref{eq:13})].
  \item {\ttfamily complex(dp), dimension(num\_sym), intent(out) :: kphase}                 \\ The k-dependent phase factors due to the translation part ${\bf v}_s$ [\ie $e^{-i{\bf k}\cdot {\bf v}_s}$ in Eq.~(\ref{eq:pw})].
  \item {\ttfamily complex(dp), dimension(dim\_basis,num\_bands), intent(out) :: Gphase\_pw}  \\ The phase factors dependent on the PW vectors [\ie $e^{-i{\bf G}_{j^\prime}\cdot {\bf v}_s}$ in Eq.~(\ref{eq:pw})].
  \item {\ttfamily integer(dp), dimension(dim\_basis,num\_bands), intent(out) :: rot\_vec\_pw}  \\ The transformation vectors of $R_s$, which send the $j$th PW to the $j^\prime$th PW [\ie ${\bf G}_{j^\prime} \equiv R_s({\bf k}+{\bf G}_j)-{\bf k}$ in Eq.~(\ref{eq:pw})].

\end{itemize}

\newpage
\lstset{language=Fortran, keywordstyle=\color{blue!70}, basicstyle=\ttfamily, frame=shadowbox}
\begin{lstlisting}
  subroutine tb_setup(WK, lattice, &
                      num_sym, det, angle, axis, tau, &
                      num_atom, atom_position, &
                      dim_basis, num_basis, angularmom, orbt, &
                      rot, SO3, SU2, &
                      kphase, rot_mat_tb)
\end{lstlisting}
\begin{itemize}
  \item {\ttfamily real(dp), dimension(3), intent(in) :: WK}                               \\ The coordinates of the k-point with respect to primitive reciprocal lattice vectors.
  \item {\ttfamily real(dp), dimension(3,3), intent(in) :: lattice(3,3)}                   \\ The primitive lattice vectors in Cartesian coordinates [\ie $({\bf t_1},{\bf t_2},{\bf t_3})$ in Eq.~(\ref{eq:14})].
  \item {\ttfamily integer, intent(in) :: num\_sym}                                        \\ The number of space-group operations $\mathcal{O}_s \equiv \{R_s|{\bf v}_s\}$ (module the integer lattice translations).

  \item {\ttfamily real(dp), dimension(num\_sym), intent(in) :: det}                       \\ The determination of the rotation part $R_s$ of space-group operations $\mathcal{O}_s$ [\ie $Det$ in Eq.~(\ref{Eq:realR})].
  \item {\ttfamily real(dp), dimension(num\_sym), intent(in) :: angle}                     \\ The rotation angle of space-group operations $\mathcal{O}_s$ [\ie $\omega$ in Eq.~(\ref{Eq:realR})].
  \item {\ttfamily real(dp), dimension(3,num\_sym), intent(in) :: axis}                    \\ The rotation axis of space-group operations $\mathcal{O}_s$ in Cartesain coordinates [\ie $\vec{n}$ in Eq.~(\ref{Eq:realR})].

  \item {\ttfamily real(dp), dimension(3,num\_sym), intent(in) :: tau}                     \\ The translation part ${\bf v}_s$ of space-group operations $\mathcal{O}_s$ with respect to primitive lattice vectors.
  \item {\ttfamily integer, intent(in) :: num\_atom}                                       \\ The number of atoms in the TB Hamiltonian.
  \item {\ttfamily real(dp), dimension(3, num\_atom), intent(in) :: atom\_position}        \\ The coordinates of atoms with respect to primitive lattice vectors [\ie $\tau_{\mu}$ in Eq.~(\ref{eq:fft})].
  \item {\ttfamily integer, intent(in) :: dim\_basis}                                      \\ 
   The reserved number of the TB basis ({\ttfamily dim\_basis = num\_basis}).
  \item {\ttfamily integer, intent(in) :: num\_basis}                                      \\ The number of orthogonal local orbitals for the k-point.
  \item {\ttfamily integer, dimensiont(num\_atom), intent(in) :: angularmom}               \\ The local orbital information on each atom. Detailed explainations can be found in Table~\ref{tab:conv1}.
  \item {\ttfamily integer, intent(in) :: orbt}                                            \\
  The convention of the local obitals on each atom. \\
  If {\ttfamily orbt = 1}, local orbitals are in the order of Table~\ref{tab:conv1}.\\
  If {\ttfamily orbt = 2}, local orbitals are in the order as implemented in Wannier90 
  \item {\ttfamily integer, dimension(3,3,num\_sym), intent(out) :: rot}                   \\ The rotation part $R_s$ of $\mathcal{O}_s$ with respect to primitive lattice vectors [\ie the matrix $Z$ in Eq.~(\ref{eq:defZ})].
  \item {\ttfamily real(dp), dimension(3,3,num\_sym), intent(out) :: SO3}                  \\ The $R_s$ given in Cartesian coordinates [\ie $R(\omega, \vec{n})$ in Eq.~(\ref{Eq:realR})].
  \item {\ttfamily complex(dp), dimension(2,2,num\_sym), intent(out) :: SU2}               \\ The $R_s$ given in spin-1/2 space [\ie $S(\omega, \vec{n})$ in Eq.~(\ref{eq:13})].
  \item {\ttfamily complex(dp), dimension(num\_sym), intent(out) :: kphase}                 \\ The k-dependent phase factors due to the translation part ${\bf v}_s$ [\ie $e^{-i(R_s {\bf k}\cdot {\bf v}_s)}$ in Eq.~(\ref{eq:9})].
  \item {\ttfamily integer(dp), dimension(dim\_basis,dim\_basis,num\_bands), intent(out) :: rot\_mat\_tb}  \\ The transformation matrices of $R_s$ in the orthogonal TB basis [\ie $\overline{V(R_s{\bf k}-{\bf k})D}$ in Eq.~(\ref{eq:9})].
\end{itemize}

\clearpage
\end{document}